\documentclass{article}
\usepackage[utf8]{inputenc}
\usepackage[top=30mm, bottom=40mm, left=25mm, right=25mm]{geometry}
\usepackage{graphicx}
\usepackage{hyperref}
\usepackage{bm}	
\usepackage{amsmath,amssymb} 	
\usepackage{empheq}
\usepackage{fancybox}

\usepackage{mathrsfs} 
\usepackage{latexsym}
\usepackage{color}
\usepackage{dsfont}
\numberwithin{equation}{section}
\definecolor{rougef}{rgb}{0.56,0,0}
\definecolor{vertf}{rgb}{0,0.5,0}
\definecolor{bleuf}{rgb}{0,0,0.8}

\usepackage{amsfonts}

\makeindex

\def\be{\begin{equation}}
\def\ee{\end{equation}}
\def\bea{\begin{eqnarray}}
\def\eea{\end{eqnarray}}
\begin{document}

{\huge{Action principles for higher and fractional spin gravities}}
\vspace{.2cm}

\begin{center}
{\textbf{C. Arias$\,{}^{a}$, 
R.~Bonezzi$\,{}^{b}$,
N.~Boulanger$\,{}^{c}$,  
E.~Sezgin$\,{}^{d}$,
P.~Sundell$\,{}^{e}$,
A.~Torres-Gomez$\,{}^{f}$}} and 
{\textbf{M.~Valenzuela$\,{}^{g}$
}}
\end{center}
\vspace{.2cm}


\begin{center}
{\em
${}^{\mbox{\footnotesize a,e,f}}$
Departamento de Ciencias Físicas, Universidad Andres Bello,
Republica 220, Santiago, Chile
~\\
\vspace{.4cm}
${}^{\mbox{\footnotesize b,c}}$
Groupe de M\'ecanique et Gravitation, Theoretical and Mathematical Physics unit,\\ 
Universit\'e de Mons--UMONS, 
20 Place du Parc, 7000 Mons, Belgium
~\\
\vspace{.4cm}
${}^{\mbox{\footnotesize d}}$
George and Cynthia Woods Mitchell Institute for Fundamental
Physics and Astronomy \\ 
Texas A\& M University, College Station,
TX 77843, USA
~\\
\vspace{.4cm}
${}^{\mbox{\footnotesize g}}$
Facultad de Ingeniería y Tecnología\\ 
Universidad San Sebastián, General Lagos 1163, Valdivia 5
110693, Chile
}
\end{center}


\begin{abstract}
We review various off-shell formulations for 
interacting higher-spin systems in dimensions 
3 and 4.
Associated with higher-spin systems in 
spacetime dimension 4 is a Chern--Simons action 
for a superconnection taking its values in a 
direct product of an infinite-dimensional 
algebra of oscillators and a Frobenius algebra. 
A crucial ingredient of the model is that it 
elevates the rigid closed and central two-form 
of 
Vasiliev's theory to a dynamical 2-form and 
doubles the higher-spin algebra, thereby 
considerably reducing the number of possible 
higher spin invariants and giving a nonzero 
effective functional on-shell. 
The two action principles we give for higher-spin systems in 3D are 
based on Chern--Simons and BF models. 
In the first case, the theory we give unifies higher-spin gauge fields
with fractional-spin fields and an internal sector. 
In particular, Newton's constant is related to the coupling 
constant of the internal sector.
In the second case, the BF action we review gives the fully
nonlinear Prokushkin--Vasiliev, bosonic equations for matter-coupled 
higher spins in 3D. We present the truncation to a single, real 
matter field relevant in the Gaberdiel-Gopakumar holographic duality. 
The link between the various actions we present is the fact that
they all borrow ingredients from Topological Field Theory. 
It has bee conjectured that there is an 
underlying and unifying
2-dimensional first-quantised description of 
the previous higher-spin models in 3D and 4D, 
in the form of a Cattaneo--Felder-like 
topological action containing fermionic fields.
\end{abstract}

\vspace*{1.5cm}
\begin{flushleft} \footnotesize
{${}^a$ \href{mailto:ce.arias@uandresbello.edu}{ce.arias@uandresbello.edu}} \hspace{.5cm}
{${}^b$ \href{mailto:}{roberto.bonezzi@umons.ac.be}}\hspace{.5cm} 
{${}^c$ \href{mailto:nicolas.boulanger@umons.ac.be}{nicolas.boulanger@umons.ac.be}}\hspace{.5cm}
{${}^d$ \href{mailto:sezgin@tamu.edu}{sezgin@tamu.edu}}\\
{${}^e$ \href{mailto:per.anders.sundell@gmail.com}{per.anders.sundell@gmail.com}}\hspace{.5cm}
{${}^f$ \href{mailto:alexander.torres.gomez@gmail.com}{alexander.torres.gomez@gmail.com}}\hspace{.5cm}
{${}^g$ \href{mailto:valenzuela.u@gmail.com}{valenzuela.u@gmail.com}}
\end{flushleft}

\setcounter{page}0
\thispagestyle{empty}

\newpage
\tableofcontents

\section{Introduction}

For our contribution to the proceedings of the 
International Workshop on Higher Spin Gauge Theories that took place 
in Singapore  on 5--7 November 2015,  
we review a number of results presented in 
Refs.~\cite{Arias:2015wha,Boulanger:2015uha,Boulanger:2015kfa,Bonezzi:2015igv} 
where off-shell formulations of nonlinear higher-spin 
systems have been found that describe interacting higher-spin 
fields in spacetime dimensions 3 and 4.

In Section \ref{sec:FCS} we first review the work~\cite{Boulanger:2015kfa} 
giving a Frobenius--Chern--Simons model for nonlinear higher-spin theory 
in 4D. This model makes use of ingredients provided in the geometrical formulation of 
higher spin gravity \cite{Sezgin:2011hq} and the action principle proposed 
in Ref.~\cite{Boulanger:2011dd}. The important new properties are
the introduction of a dynamical 2-form and the attendant phenomenon 
of the higher spin algebra doubling.  This leads to a more predictive 
power, as it restricts the possible higher spin invariant functionals.

In Section \ref{sec:FSGRA} based on Ref.~\cite{Boulanger:2015uha}, we review 
the construction of 3D higher-spin (HS) models coupled to an internal 
$U(\infty)\otimes U(\infty)$ sector and fractional-spin fields. 
The latter fields generalise the gravitini and our model can be seen as 
an extension of the Achucaro--Townsend Chern--Simons supergravity \cite{Achucarro:1987vz}.
In particular, the couping constant in the $U(\infty)\otimes U(\infty)$ internal sector
is proportional to the gravitational Newton constant.

In Section \ref{sec:PVBlencowe} based on Ref.~\cite{Bonezzi:2015igv}, 
we review the construction of an action that reproduces the fully nonlinear and bosonic 
Prokushkin--Vasiliev equations \cite{Prokushkin:1998bq}. 
The action is shown to restrict to the Blencowe action \cite{Blencowe:1988gj,Campoleoni:2010zq}
thereby reproducing the standard kinetic terms in the higher-spin sector of the action. 
We also review from Ref.~\cite{Bonezzi:2015igv}
the unfolded nonlinear equations for 3D HS fields coupled to a single real scalar field 
in the matter sector, as this is relevant in the context of the Gaberdiel--Gopakumar duality 
\cite{Gaberdiel:2010pz,Gaberdiel:2012uj}.

Finally, in Section \ref{sec:TOS} we review the works \cite{Arias:2015wha,Bonezzi:2015lfa}
where a topological open-string model of the Cattaneo--Felder type \cite{Cattaneo:2001bp}
was proposed as an underlying first-quantised model for higher-spin gauge theory.

\section{Frobenius-Chern-Simons Action for 4D Higher Spin Gravity}
\label{sec:FCS}

An outstanding problem in higher spin gravity in four (and higher) dimensions 
\cite{Vasiliev:1990en,Vasiliev:1992av,Vasiliev:2003ev} is to find 
an action principle with desirable properties. Treating this 
problem as nonlinear completion of Fronsdal kinetic terms in a 
Noether procedure approach runs  into considerable technical difficulties. 
Indeed, in the metric-like \cite{Fronsdal:1978rb} and the related frame-like  
\cite{Vasiliev:1986td,Lopatin:1987hz,Vasiliev:2001wa} approaches, 
long term efforts --- see for 
example \cite{Fradkin:1986qy,Vasilev:2011xf}
and \cite{Buchbinder:2006eq,Metsaev:2006ui,Fotopoulos:2007yq,Zinoviev:2008ck,Boulanger:2008tg,Boulanger:2011qt,Joung:2012hz,Boulanger:2012dx,Joung:2013nma} 
in the case of $AdS$ background --- has so far led to 
primarily cubic interactions.
Beyond the cubic order, the fact that higher spin 
gravity has a mass scale  set by the 
bare cosmological constant while nonabelian higher spin 
symmetries  require higher derivative vertices, lead to 
intractable abelian vertices built from 
curvatures and their higher derivatives; 
see \cite{Boulanger:2008tg} and the review \cite{Bekaert:2010hw}.

The Noether procedure approach does not exploit the fact that  
Vasiliev's equations \cite{Vasiliev:1990en,Vasiliev:1992av,Vasiliev:2003ev} 
provide a fully non-linear description of higher spin  gravity on-shell. Furthermore, 
it is background dependent procedure since it is based on perturbation around $AdS_4\,$. 
Both drawbacks can be avoided by considering covariant Hamiltonian actions from which the 
background independent full Vasiliev equations follow. 
These equations are Cartan integrable systems of differential 
forms on special noncommutative manifolds taking their values in 
associative higher spin algebras.
Treating these forms as the fundamental fields following the AKSZ approach
\cite{Alexandrov:1995kv}, one is led to a path integral formulation \cite{Boulanger:2012bj} 
based on covariant Hamiltonian actions \cite{Boulanger:2011dd} on 
noncommutative manifolds with boundaries.
The importance of the boundaries, which are absent
in the related proposals \cite{Vasiliev:1988sa,Doroud:2011xs},
is that they facilitate the deformation of the 
bulk action by boundary terms \cite{Sezgin:2011hq}
that contribute to the action but not its variation on-shell. 
The resulting higher spin amplitudes reproduce desired 
holographic correlation functions 
\cite{Colombo:2010fu,Colombo:2012jx,Didenko:2012tv},
which are suggestive of an underlying topological open string
\cite{Engquist:2005yt,Arias:2015wha,Bekaert:2015tva}. 
However, the presence of a large number of free parameters impede the 
predictive power of the model.


A more predictive model has been proposed in Ref.~\cite{Boulanger:2015kfa} 
in which the  closed and central holomorphic two form in Vasiliev equations is 
elevated to a dynamical two-form master field, and the higher spin algebra 
is necessarily extended to includ  new master one-form field.  The model also
employs an eight dimensional Frobenius algebra and provides an action which 
takes the form of a Chern--Simons term for a superconnection that accommodates 
all the master fields. For this reason the model is referred to as the  
Frobenius--Chern--Simons  (FCS) gauge theory. The construction of the model has 
been described in detail  in Ref.~\cite{Boulanger:2015kfa}. Here we shall summarize 
its salient features.


\subsection{Base manifold}


The model is formulated in terms of differential forms on the direct product space
${\cal M}_{9}= {\cal X}_{5}\times {\cal Z}_4$,  where ${\cal X}_{5}$ is a five-dimensional 
commutative manifold with boundary $ {\cal X}_{4}=\partial {\cal X}_{5}$, containing the 
original spacetime manifold ${\cal M}_4$ as a possibly open subset, and ${\cal Z}_4$ is 
a four-dimensional noncommutative space without boundary. Thus,
$\partial{\cal M}_9={\cal X}_4\times {\cal Z}_4$, where  ${\cal X}_5={\cal X}_4\times [0,\infty[\;$. 

The topology of ${\cal Z}_4$ may be chosen in a variety 
ways with nontrivial and interesting consequences to be investigated. 
In Ref.~\cite{Boulanger:2015kfa}, 
 ${\cal Z}_4$ is obtained from the  standard noncommutative ${\mathbb{C}}^4$ by choosing a real 
form and a compatible convolution formula for the star 
product and then adding points at infinity to create a 
compact noncommutative space that can be used to define 
a (graded cyclic) trace operation.
Moreover, ${\cal Z}_4$ is taken to be closed 
to avoid boundary terms, and that its structure admits 
a certain closed two-form and a global $SL(2;{\mathbb{C}})$ 
symmetry, in order to make contact with Vasiliev's theory.
To this end, one introduces canonical coordinates
$(z^\alpha,\bar z^{\dot{\alpha}})$ ($\alpha,\dot{\alpha}=1,2$) and anti-commuting
differentials $(dz^\alpha,d\bar z^{\dot{\alpha}})$ on ${\mathbb{C}}^4$.
We then consider a formally defined associative star product algebra
given by  the space $\Omega({\mathbb{C}}^4)$ of differential forms
equipped with two associative composition rules, namely the
standard graded commutative wedge product rule, denoted by juxtaposition,
and the graded noncommutative rule
\be
 f\star g= f \exp\left( - i(\overleftarrow {\partial}^{\alpha}\overrightarrow{\partial_{\alpha}}
+\overleftarrow {\bar\partial}^{\dot{\alpha}}\overrightarrow{\bar\partial}_{\dot{\alpha}})\right)\, g\ ,
\label{starZ}
\ee
The star product is thus the representation using Weyl ordering
symbols of the associative algebra of composite operators
built from anti-commuting line elements and noncommutative
coordinates with canonical commutation rules
\be 
[z^\alpha,z^\beta]_\star=-2i\epsilon^{\alpha\beta}\ ,\qquad [z^\alpha,\bar z^{{\dot{\alpha}}}]_\star=0\ ,
\qquad [\bar z^{\dot{\alpha}}, \bar z^{\dot{\beta}}]_\star=-2i\epsilon^{\dot{\alpha}\dot{\beta}}\ .
\ee
In models with four-dimensional Lorentz symmetry, it is natural to select real forms on the real slice
\be 
{\cal R}^{{\mathbb{C}}}_4 = \left\{ (z^\alpha,\bar z^{\dot{\alpha}})\,:\ (z^\alpha)^\dagger=-\bar z^{\dot{\alpha}}\ ,\quad
(\bar z^{\dot{\alpha}})^\dagger=-z^\alpha\,\right\}\, \cong \, {\mathbb{C}}^2 \times \overline {{\mathbb{C}}^2}\,\subset\, {\mathbb{C}}^4\ ,
\label{realZ}
\ee
on which $z^\alpha$ is thus a complex doublet.

In order to include Gaussian elements and distributions, it is useful 
to first introduce auxiliary integral representations of the star product \eqref{starZ}
as follows
\be 
f\star g= \int_{ {\cal R}^{\mathbb{R}}_4}  \frac{d^2\xi d^2\tilde\xi}{(2\pi)^2} 
\int_{ {\cal R}^{\mathbb{R}}_4} \frac{d^2\eta d^2\tilde\eta}{(2\pi)^2}
e^{i(\eta^\alpha \xi_\alpha+\tilde\eta^{\dot{\alpha}}\tilde\xi_{\dot{\alpha}})}
f(z+\xi,\bar z+\tilde\xi; dz,d\bar z) g(z-\eta,\bar z-\tilde \eta;dz,d\bar z) ,
\label{starZI}
\ee
where the integration domain is chosen conveniently as the real 
\be
 {\cal R}^{\mathbb{R}}_4= \left\{(\xi^\alpha,\tilde\xi^{\dot{\alpha}}): \ \xi^\alpha,\tilde\xi^{\dot{\alpha}} \in \mathbb{R}^2\right\}\cong \mathbb{R}^2\times \mathbb{R}^2
\label{domain}
\ee
The graded cyclic trace operation on  $\Omega({\cal Z}_4)$ is defined as 
\be 
{\rm STr}_{\Omega({\cal Z}_4)}\, f = \int_{{\cal R}^{\mathbb{R}}_4} f \ ,
\label{TrZ}
\ee 
which projects onto the top form in $f$.  Thus, if $f$ is a top form in $\Omega({\cal Z}_4)$ 
then its representative in $\Omega({\cal R}^{\mathbb{R}}_4)$ 
must fall off sufficiently fast at infinity for the 
integral to be convergent. 
Therefore,  ${\cal Z}_4$ must be a compact manifold obtained by 
adding points to ${\cal R}^{\mathbb{R}}_4$ at infinity to extend 
its differential Poisson algebra structure
\cite{Chu:1997ik,Beggs:2003ne,McCurdy:2008ew,McCurdy:2009xz} 
(see also \cite{Arias:2015wha,Bonezzi:2015lfa}).
This can be achieved by assuming that ${\cal Z}_4$ admits 
a Poisson structure and a compatible pre-connection. 
The latter is assumed to be trivial for simplicity.
In addition,  ${\cal Z}_4$ is required to be closed, such that
$ {\rm STr}_{\Omega({\cal Z}_4)} \, df=0$,  in order to avoid boundary 
terms from  ${\cal Z}_4$ in varying the FCS action.
Assuming that $f$ and $g$ are two smooth symbols that fall off sufficiently
fast, it follows from \eqref{starZI} that  
\be {\rm STr}_{\Omega({\cal Z}_{4})}\, f\star g=
\int_{{\cal R}^\mathbb{R}_4} f\star g=\int_{{\cal R}^\mathbb{R}_4} fg\ ,
\label{fstarg}
\ee
which is graded cyclic. As this property will be useful in analyzing boundary conditions
in the FCS model arising from the variational principle, we shall
assume that \eqref{fstarg} holds for all elements in $\Omega({\cal Z}_{4})$
including distributions.
Finally, in order to obtain Vasiliev's equations from the FCS action, 
it is assumed that  $\Omega({\cal Z}_4)$ admits global 
$SL(2;{\mathbb{C}})$ symmetry and contains the (globally defined) 
closed two-forms
\be 
j_z=-\frac{i}4 dz^\alpha dz_\alpha \kappa_z\ ,\qquad \bar j_{\bar z} =(j_z)^\dagger\ ,
\label{jz}
\ee
where the inner Klein operator
\be 
\kappa_z=2\pi \delta^2(z^\alpha)\ .\label{kappayz}
\ee

A choice of topology that satisfies all of the requirements as stated 
above iis given by\footnote{The manifolds $S^4$ and $S^3\times S^1$ 
do not admit $j$, while $T^2 \times T^2$ breaks global 
$SL(2;{\mathbb{C}})$ symmetry.}
\be \Omega({\cal Z}_4)= \bigoplus_{m,\bar m=0,1}
(\Omega(S^2)\star (j_z)^{\star m} )\otimes (\Omega(S^2)\star ({\bar j}_{\bar z})^{\star {\bar m}})\ ,\ee
where $\Omega(S^2)$ consists of globally defined 
forms on $S^2$ with Poisson structure obtained 
by extending the Poisson structure of \eqref{starZ}
to the point at $\infty$.
At this point, the resulting Poisson bivector and 
all its derivatives vanish.
Hence, provided it is possible to exchange the
order of differentiation and summation in \eqref{starZ} 
and using the fact that increasing number of derivatives 
of a form that falls off yields forms that fall off even 
faster, it follows that if $f,g\in \Omega(S^2)$ then $ (f\star g)|_\infty=f|_\infty g|_\infty$
i.e. the point at infinity is a commuting  point of $\Omega(S^2)$.
In other words, one is working with a topological two-sphere equipped with a differential Poisson
algebra with trivial pre-connection.
Moreover, in order for the elements in $\Omega({\cal Z}_4)$
to have finite traces,  it is assumed that  the top forms on each
two-sphere fall off sufficiently fast at infinity working
in the original $\mathbb{R}^2\times \mathbb{R}^2$ coordinate chart.
For this fall-off condition to be embeddable in a differential
star product algebra, also the forms in lower degrees must
fall off appropriately at infinity. 
In particular, the only forms that can have finite values at infinity
are the zero-forms. 
Thus, in effect, one has 
\be 
{\cal Z}_4=S^2 \times S^2\ ,
\ee
by assuming boundary conditions at the commuting points 
at infinity and allowing for delta function distributions 
at the origins so as to create a space of forms that is closed 
under exterior differentiation and star products, and has a
space of top forms with finite traces that vanish for
exact elements and obey \eqref{fstarg}.
For this to hold true it is important that the delta function
$\kappa_z$ always appears together with line elements in the 
combination $j_z$ given in \eqref{jz}, which obeys $j_z\star j_z=0$, 
whereas the inclusion of $\kappa_z$ into the algebra would 
require the inclusion of $j_z\star \kappa_z$ as well
which is not integrable.
%


\subsection{Superconnection and Frobenius-Chern-Simons action}


The construction of the FCS action  employs an eight-dimensional,  3-graded Frobenius algebra 
\be {\cal F}={\cal F}^{(-1)}\oplus{\cal F}^{(0)}\oplus {\cal F}^{(+1)}\ ,
\qquad (e_{ij}, h\, e_{ij})\in {\cal F}^{(i-j)}\ ,\ee
where $e_{ij}$ is the $2\times 2$ matrix whose only non-vanishing entry
is a $1$ at the $i$th row and $j$th column, and $h$ is a Klein 
element satisfying $[h,e_{11}]=0=[h,e_{22}]\ , \{h,e_{12}\}=0 = \{h,e_{12}\}\,$.
The coordinate-like master fields are assembled into
\be
X = \sum _{i,j} X^{ij} e_{ij} = \begin{pmatrix}
A & B \\ \widetilde B & \widetilde A 
\end{pmatrix}\ ,
\label{XX}
\ee
and the momentum-like master fields, which will play the role of Lagrange multipliers, into 
\be
P = \sum _{i,j} P^{ij} e_{ij} = \begin{pmatrix}
V & U \\   \widetilde U & \widetilde V
\end{pmatrix}\ .
\label{PP}
\ee
The above master fields are duality extended in the sense that they are formal sums of 
forms with degrees 
\begin{align}
\begin{split}
& {\rm deg} (B,A,\widetilde A,\widetilde B) = 
\left\{(2n,1+2n,1+2n,2+2n)\right\}_{n=0,1,2,3}\ ,
\\
& {\rm deg} 
(\widetilde U,V,\widetilde V,U)=\left\{(8-2n,7-2n,7-2n,6-2n)
\right\}_{n=0,1,2,3}\ .
\end{split}    \label{contentDE}
\end{align}
One then proceeds by defining a superconnection and superdifferential \cite{Quillen198589} as
\be 
Z = hX+ P\ ,
\label{ZZ}
\ee
respectively, which are thus objects with odd total degree  given by form degree plus Frobenius 3-degree. 
The space $\Omega({\cal M}_9)$ of differential forms is equipped with two associative 
composition rules, namely the standard graded commutative wedge product rule, 
denoted by juxtaposition, and Weyl ordered star products of functions. 
The need for Weyl ordering is due to the property \eqref{fstarg} that 
is crucial for the boundary conditions to make sense in a noncommutative set up.
Once one has put any star-product expression in its factorized form in $Y$ and $Z$, 
say $F(Y) \star G(Z)$  equal to $F(Y) G(Z)$ in Weyl order, 
one makes assumptions about the functional classes to which $F(Y)$ and $G(Z)$ belong. 
See Section 3.7 of \cite{Iazeolla:2011cb} for details. 

For reasons explained in detail in \cite{Boulanger:2015kfa}, the action function is 
expressed in terms of globally defined configurations in
\be 
{\cal E}=
\Omega({\cal X}_5)\otimes \frac12(1+\pi \bar\pi) \left[ \Omega({\cal Z}_4) \otimes  
{\cal A} \otimes \tfrac12(1+k\star\bar k)\right] \ ,
\ee
where 
\be
{\cal A} = {\cal F} \otimes {\cal W}_0 \otimes {\cal K}\ .
\ee
Furthermore,  ${\cal W}_0$ is an extended Weyl algebra 
\be 
{\cal W}_0  =\bigoplus_{r, \bar r =0,1} 
{\rm Aq}(2) \star (\kappa_y)^{\star r} \star (\bar \kappa_{\bar y})^{\star \bar r}
\label{calAmm}
\ee
where $Aq(2)$ consists of star polynomials in two complex doublets $(y^\alpha,\bar y^{\dot{\alpha}})$, 
$\alpha,\dot{\alpha}=1,2$, obeying the oscillator algebra
\be
[ y^\alpha, y^\beta]_\star= 2i\epsilon^{\alpha\beta}\ ,\quad [ y^\alpha, \bar y^{\dot{\beta}}]_\star=0\ ,
\quad
[ {\bar{y}}^{\dot{\alpha}},{\bar{y}}^{\dot{\beta}}]_\star = 2 i\epsilon^{\dot{\alpha}\dot{\beta}}\ .
\label{oscalg}
\ee
The inner Klein operators in Weyl order are defined as
\be \kappa_y:= 2\pi \delta^2(y)\ ,\qquad
\bar\kappa_{\bar y}:= 2\pi \delta^2(\bar y)\ ,
\ee
so that $P\in {\cal W}_0$ obey 
\be 
\kappa_y\star P\star \kappa_y=\pi_y(P)\ ,\qquad
\bar\kappa_{\bar y}\star P\star \bar\kappa_{\bar y}=\bar\pi_{\bar y}(P)\ ,
\ee
where $\pi_y$ and $\bar\pi_{\bar y}$ are inner automorphisms  
whose action in Weyl order is given by
\be 
\pi_y (y) =-y\ ,\qquad  \bar\pi_{\bar{y}} (\bar{y}) = -\bar{y}\ ,
\label{py}
\ee
leaving intact all the Klein operators.  Thus, the generic elements of ${\cal W}_0$ 
is of the form
\be 
P = \sum_{r,\bar r=0,1} P^{r,\bar r}_{\alpha(n),\dot\alpha (\bar n)}  
(\kappa_y)^{\star r}\star (\bar\kappa_{\bar y})^{\star \bar r} 
y^{(\alpha_1}\star \cdots\star y^{\alpha_n)}\star  
\bar y^{(\dot{\alpha}_1}\star\cdots\star \bar y^{\dot{\alpha}_{\bar n})}\ .
\label{exp}
\ee

Turning to ${\cal K}$, it is the collection of outer Klein operators
\be
{\cal K}=\{ 1, k, {\bar k} , k \star {\bar k} \}\ .
\label{defcalA}
\ee
where for $f\in {\cal E}$, the adjoint action 
\be
k\star f\star k = \pi (f)\ ,\qquad \bar k \star f \star \bar k = \bar \pi (f)\ ,
\ee
and the outer automorphisms $\bar\pi$ and $\bar\pi$ with the only nontrivial actions
\be
\pi (y,z) = (-y,-z)\ ,\qquad \bar\pi (\bar y, \bar z) = (-\bar y,-\bar z)\ .
\ee
Employing the ingredients summarized above, the following action has been 
proposed \cite{Boulanger:2015kfa}
\be
S  =  \int_{{\cal M}_9} {\rm Tr}_{{\cal A}}
\left(\tfrac{1}{2}  \,Z\star q Z+\tfrac{1}{3}\, Z\star Z\star Z\right)
-\frac{1}{4}\, \int_{\partial{\cal M}_9} {\rm Tr}_{{\cal A}} \,[h\pi_h(Z) \star Z]\ ,
\label{ss1}
\ee
where $\pi_h$ is the automorphism sending $h$ to $-h\,$, and
\be
q := hd\ .
\ee
Keeping in mind that $f \in {\cal E}$, the operation ${\rm Tr}_{\cal A}$ is defined as
\be 
S= \int_{{\cal M}_9} {\rm Tr}_{\cal A}\ f := \int_{{\cal X}_5}{\rm STr}_{\Omega({\cal Z}_4)}
\, {\rm Tr}_{\cal F}\, {\rm Tr}_{{\cal W}_0}\, {\rm Tr}_{\cal K} f\ ,
\ee
where ${\rm STr}_{\Omega({\cal Z}_4)}$ is defined in \eqref{TrZ}, and  the remaining trace 
operations are defined as
\bea 
&& {\rm Tr}_{{\cal F}} \sum_{i,j} e_{ij} M^{ij}(h)=M^{11}(0)+
M^{22}(0)\ , 
\label{tt1}\\
&&  {\rm Tr}_{{\cal W}_0} P = P^{1,1}(0,0)\ ,  
\label{tt2}\\
&& {\rm Tr}_{\cal K} f=f|_{k=0=\bar k}\ ,
\label{tF}
\eea
with $P\in {\cal W}_0$ from \eqref{exp} which furnishes the definition of $P^{1,1}(0,0)$. 
Defining $ {\rm Tr}_{{\cal E}} f  := \int_{{\cal M}_9} {\rm Tr}_{\cal A}\ f$,  it can be 
shown that  ${\rm Tr}_{{\cal E}}\ f\star g = {\rm Tr}_{{\cal E}}\  g\star f$ \cite{Boulanger:2015kfa}.

The total variation of the FCS action gives
\be 
\delta S = \int_{{\cal M}_9} 
{\rm Tr}_{\cal A}\ 
\delta Z\star R +\frac12 \oint_{{\cal M}_9} 
{\rm Tr}_{\cal A}\ h\,\delta Z\star \left( Z+ \pi_h(Z)\right)\ ,
\label{deltaFCS}
\ee
where the Cartan curvature
\be
R := qZ + Z\star Z\ . 
\label{R}
\ee
Thus, imposing the boundary condition
\be
\left( Z+ \pi_h(Z)\right)\big\vert_{\partial{\cal M}_9}=0\ ,
\ee
one has the equation of motion $R=0\,$. This equation is Cartan integrable, 
hence gauge invariant, with transformations
\be
 \delta  Z =  {q}   \theta +[ Z, \theta]_\star\ ,\qquad
\delta  R=[ R, \theta]_\star\ .
\label{ZV}
\ee
One the other hand, the requirement of gauge invariance of the action gives the 
following boundary conditions
\begin{equation}
\left(\theta - \pi_h(\theta)\right)\vert_{\partial{\cal M}_9} = 0 \ .
\label{boundaryfortheta}
\end{equation}
In obtaining this result, the property \eqref{fstarg} plays an important role. 
Setting aside nontrivial flat connections due to the noncommutativity 
of the base manifold, $Z$ can be given on-shell in terms of a gauge function 
$L$ (which contains forms in different degrees) and a zero-form 
integration constant $C$, viz.
\be 
Z= L^{\star(-1)}\star (q+C)\star L\ ,\qquad qC=C\star C=0\ ,
\label{C&L}
\ee
where the algebraic condition on $C$ is a consequence of the fact that the form 
content is as given in \eqref{contentDE}.

The superconnection $Z$ is assumed to be globally defined. However, if it is rather 
given by a set of representatives defined locally on charts that cover ${\cal M}_9$, 
the appropriate global definition of the action is described in Ref.~\cite{Boulanger:2015kfa}. 
In doing so, 
it proves convenient to write the action \eqref{ss1} in terms of  master fields $(X,P)$ 
defined via $Z=hX+P$. Thanks to the boundary term in \eqref{ss1}, 
one finds \cite{Boulanger:2015kfa}
\be
S= \int_{{\cal M}_9} {\rm Tr}_{{\cal A}} \, \left(P \star F^X
+\tfrac13 \,P\star P\star P\right) \ ,  
\label{ss2}
\ee
where $F^X := dX +hXh\star X$. The general variation of this action \eqref{ss1}  reads
\be
\delta S  =  \int_{{\cal M}_9} 
{\rm Tr}_{\cal A}\  \left( \delta X \star R^P h
+ \delta P \star R^X +{d}(\delta X\star P)\right)\ ,
\label{total}
\ee
where the total derivatives cancel between neighboring patches 
(in the interior of ${\cal M}_9$)
since $\delta X$ and $P$ belong to sections. 
Writing $R=R^X +R^P\,$ where
\begin{equation}
R^X := F^X + P\star P\ ,\qquad R^P := qP+hX\star P +P\star hX\ , 
\end{equation}
then on shell we have $R^X=0$ and $R^P=0\,$, and we are left with 
\be 
\delta S = \int_{{\cal M}_9} {\rm Tr}_{\cal A}\ 
d(\delta X\star P) = \oint_{\partial{\cal M}_9}{\rm Tr}_{\cal A}\ \delta X\star P\ .
\ee
Using crucially the property \eqref{fstarg} to  replace the star product in 
$\Omega({\cal Z}_4)$ by the classical product (keeping in mind that $j_z\star j_z=0$),
the variation becomes
\be 
\delta S =\oint_{\partial{\cal M}_9}{\rm Tr}_{\cal A}\ \delta X
\star_{\cal A}\ P\ .
\ee
Hence, if $X$ is free to fluctuate at $\partial{\cal M}_9$,
it follows from the variational principle that
\be 
P|_{\partial {\cal M}_9}= 0\ .
\ee
Finally, while action is invariant under the gauge transformations with parameters $\epsilon^X$,
it transforms into a total derivative under transformations with parameters $\epsilon^P$, viz.
\be 
\delta_{\epsilon^P} S =\int_{{\cal M}_9} {\rm Tr}_{\cal A}\ 
{d}\left(\epsilon^P\star R^X\right)\ ,
\ee
that vanishes provided that $\epsilon^P$ belongs to the same section as $P$, and
\be
\epsilon^P|_{\partial{\cal M}_9}=0\ 
\ee
that is indeed equivalent to \eqref{boundaryfortheta}.

Using $Z=hX+P$, the gauge transformations \eqref{ZV} read 
\begin{align}
    \begin{split}
\delta X =& {d} \epsilon^X + X\star \epsilon^X - h {\epsilon}^X h  \star X 
+ hPh\star \epsilon^P -\epsilon^P \star P\ ,
\\
\delta P =& {d} \epsilon^P + h X h\star  \epsilon^P - h \epsilon^P h \star X+ [P,\epsilon^X]_\star  \ .
\label{xpt}
\end{split}
\end{align}
%


\subsection{Component formulation}
\label{Sec:4.3}

The action \eqref{ss1}, upon using  the definitions \eqref{XX}, \eqref{PP} and \eqref{ZZ}, takes the form
\bea
S &=&  \int_{{\cal M}_9} {\rm Tr}_{{\cal W}\otimes {\cal K}}\Big[
\widetilde U \star DB  + V\star \left( F -  B \star\widetilde B 
+\tfrac{1}{3}\, V^{\star2}+ U \star\widetilde U \right)
\nonumber\\
&&  \ \ \ \ \ \ \ +U \star \widetilde D \widetilde B
+ \widetilde V\star \left( \widetilde F - \widetilde B\star B + \tfrac{1}{3}\, {\widetilde V}^{\star2} +  
\widetilde U\star U \right) \Big]\ ,
\label{SH}
\eea
where
\begin{align}
\begin{split}
& F :={d} A+A\star A\ ,\qquad\qquad\qquad\quad 
\widetilde F:= d\widetilde A +\widetilde A\star \widetilde A\ ,
\\ 
& DB := {d} B + A\star B - B\star \widetilde A \ ,\qquad 
\widetilde D \widetilde B :={d} \widetilde B
+\widetilde A\star \widetilde B-\widetilde B\star A\ ,
\\ 
& D U :={d} U+A\star U-U\star \widetilde A\ ,\qquad 
\widetilde D \widetilde U :={d} \widetilde U+
\widetilde A\star \widetilde U-\widetilde U\star A\ ,
\\ 
& D V :={d} V+A\star V+V\star  A\ ,\qquad 
\widetilde D \widetilde V :={d}\widetilde V+
\widetilde A\star \widetilde V+\widetilde V\star \widetilde A\ .
\end{split}    
\end{align}
The bulk equations of motion, which amount to vanishing Cartan curvatures, read
\begin{align}
\begin{split}
& F-B\star\widetilde B  + V\star V+U\star \widetilde U=0
\ ,\qquad DB +V\star U + U\star \widetilde V=0\ ,
\\
& \widetilde F - \widetilde B\star B +
\widetilde V \star \widetilde V+\widetilde U \star U=0\ ,
\qquad  
\widetilde D \widetilde B + \widetilde V \star \widetilde U +\widetilde U\star V=0\ ,
\\
& D U +B\star \widetilde V+ V\star B=0 \ ,\qquad DV+B\star \widetilde U-U\star \widetilde B=0\ ,
\\
& \widetilde D\widetilde U +\widetilde B\star V+\widetilde V\star \widetilde B=0\ ,\qquad  
\widetilde D \widetilde V+\widetilde B\star U-\widetilde U\star B=0\ .
\label{cc2}
\end{split}    
\end{align}
The gauge parameter can be written  as $\theta= \epsilon^X + h\epsilon^P$, where 
\be
\epsilon^X= \begin{pmatrix}   \epsilon & \eta \\ \widetilde\eta & \widetilde\epsilon \end{pmatrix}\ ,\qquad 
\epsilon^P= \begin{pmatrix}   \epsilon^V & \eta^U \\ \eta^{\widetilde U} &\epsilon^{\widetilde V} \end{pmatrix}\ .
\label{expbis}
\ee
The transformation rules for the component fields 
can be readily obtained from \eqref{xpt} by using the definitions \eqref{XX} and \eqref{PP}. 
Thus, on $\partial{\cal M}_9$, where $(U,\widetilde U;V,\widetilde V)$
vanish, one finds 
\begin{align}
\begin{split}
& F - B\star\widetilde B= 0\ ,\qquad D B= 0\ ,
\\
&
\widetilde F - \widetilde B\star B= 0\ ,\qquad \widetilde D \widetilde B= 0\ ,\label{E2}
\end{split}    
\end{align}
which is the desired modification of Vasiliev's original system \cite{Vasiliev:1990en,Vasiliev:1992av}. 
Alternatively, going to the basis 
\be
\widetilde A = W+K\ ,  \qquad A= W-K\ ,
\label{defW}
\ee
the equations of motion on $\partial{\cal M}_9$ read
\begin{align}
\begin{split}
& F_W + K\star K - \frac12\{B,\widetilde B\}_\star= 0\ ,\qquad
D_W  K - \frac12[\widetilde B,B]_\star=0\ ,
\\
& D_W  B-\{K,B\}_\star=0\ ,\qquad
D_W \widetilde B+\{K,\widetilde B\}_\star= 0\ ,
\label{eom2}
\end{split}    
\end{align}
where we have defined $D_W f={d}f + W \star f-(-1)^{\rm deg(f)}f\star W$ 
and $F_W={d}W+W^2$.
Since $\epsilon^P\vert_{\partial{\cal M}_9} = 0$, recalling the notation \eqref{expbis}, and 
splitting the gauge parameters $(\epsilon,\widetilde\epsilon)$ as
\be  \epsilon = \alpha -\beta\ ,\qquad \widetilde\epsilon = \alpha +\beta\ , \ee
the gauge transformations under which the field equations \eqref{eom2} are invariant can 
be written as 
\begin{align}
\begin{split}
\delta W &=  D_W \alpha +[K,\beta]_\star +\tfrac{1}{2}\, \{ \widetilde \eta , B \}_\star + \tfrac{1}{2}\,
\{ \widetilde B,\eta \}_\star\ ,
\\
\delta K &= D_W\beta +[K,\alpha]_\star +\tfrac{1}{2}\,\,[\widetilde\eta,B]_\star 
+ \tfrac{1}{2}\, [\widetilde B,\eta]_\star\ ,
\\
\delta B &= D_W\eta +[B,\alpha]_\star-[K,\eta]_\star  + \{ B,\beta \}_\star\ ,
\\
\delta \widetilde B &= D_W\widetilde \eta + [\widetilde B,\alpha]_\star 
+ [K, \widetilde\eta]_\star  - \{ \widetilde B , \beta \}_\star \ .
\label{dt4}
\end{split}    
\end{align}
%

\subsection{Comparison with the duality extended Vasiliev system}

In Ref.~\cite{Boulanger:2015kfa} it has been shown that bay taking $K=0$ and choosing 
${ \widetilde B }$ appropriately,  the equations of motion on $\partial{\cal M}_9$ given 
in \eqref{eom2} take the form
\begin{align}
\begin{split}
&F_W  -{\cal V} \star J + \overline {\cal V} \star {\overline J} 
+ {\cal U} _0\star J \star {\overline J} 
+  {\cal U} _1\star J_{[2]}+{\cal U} _2\star J_{[4]}=0\ ,
\\
&D_W B = 0\ ,
\end{split}
\label{EVS1}    
\end{align}
where
\be
J:=-\frac{i}{8} dz^\alpha dz_\alpha \,\kappa_z\star \kappa_y\star k  \star  (1+k \star \bar k ) \ ,\qquad\quad 
\overline J:=J^\dagger\ ,
\label{J1}
\ee
with $j_z$ defined in \eqref{jz}. Furthermore, $J_{[2]}$  and $J_{[4]}$ are 
forms that belong to the de Rham cohomology on ${\cal X}_4$, and 
$({\cal V}, \bar{\cal V},U_0, U_1, U_2)$ are star polynomial function of the 
form $f(B)= \sum_{n=0}^\infty f_n B^{\star(n+1)}$.

It is interesting to compare this system with Vasiliev's recently proposed extended system 
\cite{Vasiliev:2015mka}, adapted to our notation, given by
\begin{align}
\begin{split}
&F_W  -{\cal V} \star J + \overline {\cal V} \star {\overline J} 
+  {\cal U} _0\star J \star {\overline J} +g  J \star {\overline J}
+  {\cal L}_{[2]}+{\cal L}_{[4]}=0\ ,
\\
&D_W B = 0\ ,
\end{split}
\label{EVS2}    
\end{align}
where ${\cal L}_{[2]}$ and ${\cal L}_{[4]}$ are two new dynamical 
fields, referred to as Lagrangian forms, given by globally defined
central and closed elements of degrees two and four, respectively.
As far as the local dynamics is concerned, the two systems
are equivalent in form degrees zero and one, since one can always 
choose a representative for ${\cal L}_{[2]}$ that vanishes in a given 
coordinate chart.
In higher form degrees, the duality extended Vasiliev system contains 
the term $g J\star \overline J$ and the Lagrangian forms, which are
not present in the FCS system\footnote{
Whether such coupling can be obtained either by expanding $B$
around a constant background value or allowing the dependence 
of $\widetilde B$ on $B$ to contain a simple pole, remains to be seen.}.
In Ref.~\cite{Vasiliev:2015mka}, the integral $\oint {\cal L}_{[2]}$ has 
been interpreted as a black hole charge, as has been substantiated 
black hole solution \cite{Didenko:2008va}.
As for the integral of ${\cal L}_{[4]}$ over spacetime, 
it has been proposed \cite{Vasiliev:2015mka} as the generating
functional of correlators within the context of holography
\footnote{Another proposal for the black hole entropy and generating 
functional of correlators in higher spin gravity has been made 
in Ref.~\cite{Sezgin:2011hq}.}.
An important open 
problem in this  framework is how to account for loop corrections. 
It has been suggested that the quantum mechanical effects may emerge from 
classical dynamics in an infinite dimensional space that has enough room 
to describe all multiparticle states in the system \cite{Vasiliev:2012vf}.
If true, this would be a drastically new way of looking at quantum gravity.
The tests of these proposals remain to be seen. 

In the approach of Ref.~\cite{Boulanger:2015kfa} a path integral formulation  of 
the FCS model is proposed along the same lines as the AKSZ construction  of 
Ref.~\cite{Boulanger:2012bj}  within the geometric framework of Ref.~\cite{Sezgin:2011hq}. 
In this approach, the  terms proportional to the closed and central elements in \eqref{EVS1}, 
which are similar to the Lagrangian form terms in \eqref{EVS2} but play a different role, 
as the computation of the  effective action proceeds in this case by means of path integral 
quantization rules which  necessarily involves the FCS action itself. The advantage of this 
approach is the availability  of path integral formulation for quantization. The computation 
of quantum effects are left to  future work but an outline of the the role of certain topological 
invariants in the construction  of the on-shell effective action is given in Ref.~\cite{Boulanger:2015kfa}, 
which we summarize below. 


\subsection{On-shell actions from topological invariants}


Starting from an AKSZ path integral on ${\cal M}_9 = 
[0,\infty[\, \times {\cal X}_4\times {\cal Z}_4$,
where all fields vanish at $\{\infty\}\times {\cal X}_4\times {\cal Z}_4$ and
in addition $P|_{\{0\}\times {\cal X}_4\times {\cal Z}_4}=0$,
as required by the Batalin--Vilkovisky master equation,
one finds that $S_{\rm H}$ vanishes on-shell.
Following \cite{Sezgin:2011hq}, one may generate 
an on-shell action by adding to $S_{\rm H}$ a globally defined boundary 
term $S_{\rm top}=\oint_{\partial{\cal M}_9} {\cal V}(X,dX)$,
whose total variation vanishes off-shell, 
i.e. $S_{\rm top}$ is a topological invariant.
by its evaluation \cite{Vasiliev:2015mka} on the Didenko--Vasiliev 
Assuming that $S_{\rm top}$ does not affect the boundary
condition on $P$ nor the equations of motion,
one may argue that the on-shell action
is given by $S_{\rm top}$.

Aspects of topological invariants for a general structure group are discussed in 
Ref.~\cite{Boulanger:2015kfa}, where it is also shown that taking it to be generated 
by $\alpha$-transformation displayed in \eqref{dt4}, one has the invariants
\be 
S_{\rm top}[W,K]=\sum_{p=0}^2\sum_{n=1}^{p+2} \oint_{{\cal X}_{2p}\times
{\cal Z}_4}  \beta_{n,p}\left(\tfrac{d}{dt}\right){\rm Tr}_{{\cal W}_0
\otimes {\cal K}} \left(F_{W_t}\right)^{\star n}\Big\vert_{t=0}\ .
\label{Stop}
\ee
where ${\cal X}_{2p}\subset {\cal X}_4$  are closed subsets of dimension 
$2p$ for $p=0,1,2$; 
\be 
W_t= W+ t K\ ,\qquad F_{W_t}=F_W+ t D_W K+ t^2 K\star K\ ,
\ee
and $\beta_{n,p}$ are linear differential operators of order $(2n-1)$ in $d/dt$ 
with constant coefficients.Thus, there are $2,3,4$ invariants for $p=0,1,2$, respectively. 
The on-shell value of $S_{\rm top}[W,K]$ is built out of integrals of
traces of $B\star \widetilde B$, $\widetilde B\star B$ and $K\star K$ 
forming a finite set of invariants. The observables are invariant off shell 
under gauge transformations  with parameter $\alpha$, and on shell using 
parameters $(\beta,\eta,\widetilde \eta)$. In the semi-classical limit, one 
has the partition function \cite{Boulanger:2015kfa}
\be 
Z_{\rm FCS}=\sum_{\rm saddles} {\cal N} e^{iS_{\rm top}}\ .
\ee

\subsection{Linearized Fluctuations}
\label{sec:Boundary}

The theory on the boundary of ${\cal M}_9$ admits the vacuum solutions
\be \widetilde B^{(0)}=I\ ,\qquad W^{(0)}=L^{-1}\star dL\ ,\qquad
K^{(0)}=0\ ,\qquad B^{(0)}=0\ ,\label{vac}\ee
where $L$ is a gauge function (consisting of forms) and 
$I$ is a closed and central element on $\partial{\cal M}_9$.
In particular, to describe Vasiliev's phase of the theory, it is assumeed that
\be 
I=J_{\cal X}+e^{i\theta_0}J-e^{-i\theta_0}\overline J\ ,
\ee
where $J_{\cal X}$ is a closed a central element  
on ${\cal X}_4$, $J$ and $\overline J$ are the 
closed and central elements on ${\cal Z}_4$ defined 
in \eqref{J1}, and $\theta_0$ is an arbitrary real constant.
The fluctuations in the boundary fields can be expanded as
\be
(W-W^{(0)},B,K,\widetilde B-\widetilde B^{(0)})=\sum_{n\geqslant 1}
( W^{(n)},B^{(n)},K^{(n)},\widetilde B^{(n)})\ .
\ee
At the first order, the equations of motion \eqref{eom2} read
\begin{align}
\begin{split}
& D^{(0)} W^{(1)}-\tfrac12\{I, B^{(1)}\}_\star=0\ ,\qquad
D^{(0)} K^{(1)} =0\ , 
\\
&D^{(0)} B^{(1)}=0\ ,\qquad D^{(0)} \widetilde B^{(1)}
+\{I, K^{(1)}\}_\star=0\ ,
\label{abe}
\end{split}
\end{align}
and the abelian gauge transformations following from \eqref{dt4} are given by
\be
\delta W^{(1)}=D^{(0)} \alpha^{(1)}+\tfrac{1}{2}\,\{I, \eta^{(1)}\}_\star\ ,\qquad
\delta B^{(1)}=D^{(0)} \eta^{(1)}\ ,\ee\be
\delta K^{(1)}=D^{(0)} \beta^{(1)} \ ,\qquad
\delta \widetilde B^{(1)}=D^{(0)} \widetilde\eta^{(1)} 
- \{ I, \beta^{(1)} \}_\star\ .
\ee

Expressing the first order fluctuations as 
\be
(W^{(1)},B^{(1)},K^{(1)}, {\widetilde B}^{(1)})=
L^{-1}\star (W^{(1)\prime},B^{(1)\prime},K^{(1)\prime},{\widetilde B}^{(1)\prime})\star L\ ,
\ee
where $L$ is a gauge function and the primed fields are independent of $x$, the 
linearized equations \eqref{abe} are solved by \cite{Boulanger:2015kfa}
\bea
B^{(1)\prime}&=&B^{(1)\prime}_{[0]}+{\rm d}\rho_v B^{(1)\prime}\ ,\\[5pt]
W^{(1)\prime}&=&
d\rho_v W^{(1)\prime}-\tfrac{1}{2}\, (d\rho_v-1) \left(\{\rho_v I,
B_{[0]}^{(1)\prime} \}_\star +\{I ,\rho_v B^{(1)\prime}\}_\star\right)\ ,\\[5pt] 
K^{(1)\prime}&=&d\rho_v K^{(1)\prime}\ ,\\[5pt]
\widetilde B^{(1)\prime}
&=&d\rho_v\widetilde B^{(1)\prime}+(d\rho_v-1)\{I,\rho_v K^{(1)\prime}\}_\star\ ,
\eea
where  $B^{(1)\prime}_{[0]}$ the zero-form integration constant which harbours 
the local degrees of freedom of the system and the homotopy contractor $\rho_v$, with 
the convenient choice of $v=z^\alpha\partial_\alpha$ is defined by
\be
\rho_v f(Z,Y,dZ) = Z^{\underline{\alpha}} \frac{\partial}{\partial dZ^{\underline{\alpha}}}\, 
\int_0^1 dt \frac{1}{t} f(tZ,Y,tdZ)\ .
\ee
The connection $W^{(1)}_{[1]}$ consists of a pure gauge 
solution, as its gauge function and gauge parameter belong to the 
same spaces, plus a a set particular solutions that carrying the 
aforementioned local massless degrees of freedom.

The fields $K^{(1)}$ and $\widetilde B^{(1)\prime}$, on the other hand, 
may introduce new topological degrees of freedom arising in
cohomological spaces given by spaces of gauge functions over
the spaces of gauge parameters. 

In particular, $\widetilde B^{(1)\prime}_{[2]}$ contains moduli associated 
to the gauge function $\rho_v \widetilde B^{(1)\prime}_{[2]}=
\rho_v (e^{i\theta_0}J-e^{-i\theta_0}\bar J)$, 
as $d\rho_v (e^{i\theta_0}J-e^{-i\theta_0}\bar J)=
e^{i\theta_0}J-e^{-i\theta_0}\bar J$ belongs to an admissible section 
for $\widetilde B^{(1)\prime}_{[2]}$ while 
$\rho_v (e^{i\theta_0}J-e^{-i\theta_0}\bar J)$ does not belong 
to an admissible section for $\widetilde \eta^{(1)}$. In more detail it is shown 
in Ref.~\cite{Boulanger:2015kfa} that the moduli of $\widetilde B^{(1)\prime}_{[2]}$ can be associated to modes that blow up at  infinity, i.e. at the 
commutative point of ${\cal Z}_4$.

Going to more general backgrounds for ${\cal M}_9$, 
it follows from the fact that the fields $K$, $B$ 
and $\widetilde B$ belong to sections of the 
structure group that they can contain topological 
degrees of freedom provided that there are matching
elements in the de Rham cohomology, whose r\^ole 
remains to be investigated further. Likewise, going to 
higher 
order in perturbation theory, the moduli of $\widetilde B$ 
will generate interaction
terms which are expected to have important consequences in 
the perturbative 
expansion of the theory and the  computation of the 
correlation functions.

The above linearization suffices to show that the perturbative degrees of freedom
of the system are contained in the initial data for the Weyl zero-form. 
However, in order obtain Fronsdal field equations one has 
switch from Weyl order to normal order and perform a change of
gauge in order to make direct contact with Vasiliev's 
original perturbative expansion 
(in which $z^\alpha A_\alpha=0$ in normal order), 
which complies with the Central On Mass Shell Theorem 
(COMST).
It is important that despite the fact that the FCS model 
is formulated in the Weyl order, for reasons explained 
in Section 4, its physical spectrum agrees with the 
Vasiliev theory, and hence its perturbative expansion 
should obey the COMST as well.
Although a naive transformation of the perturbatively defined
master fields from normal to Weyl order 
is known to produce 
singularities \cite{Vasiliev:2015wma}\footnote{For a 
general discussion of ordering schemes 
and maps between them, 
see \emph{e.g.} \cite{zachos2005quantum}.},
the FCS master fields belong to an extended class of 
symbols, including inner Klein operators, which yields 
a well-defined perturbation theory in a specific 
holomorphic 
gauge (defined by $z^\alpha A_\alpha=0$ in Weyl order). 
Indeed, working with definite boundary conditions 
(corresponding to generalized Type D 
solutions \cite{Iazeolla:2007wt}), 
the resulting linearized fields can be mapped to 
Vasiliev's basis.
We plan to examine whether this remains the case for 
more general boundary
conditions and to higher orders in the perturbative 
expansion.

%

\section{Fractional spin gravity theory}\label{sec:FSGRA}
%

In $2+1$ dimensions Vasiliev's higher-spin gravity, or more specifically the 
Prokushkin--Vasiliev model \cite{Prokushkin:1998bq}, 
admits truncation to Chern--Simons higher spin gravity 
\cite{Blencowe:1988gj,Vasiliev:1989re}. 
In Ref. \cite{Vasiliev:1989re} the gauge connection is valued in a higher spin 
algebra that consists of monomials of Wigner-Heisenberg deformed oscillator operators. 
Monomials of the same order transform in spinor-tensorial representations of the $AdS_3$ algebra, 
with arbitrary half-integer spins (which includes integer and half-an-integer values), 
and the correspondent fields have standard boson or fermion statistics. 
However, in $2+1$ dimensional spacetimes the representations of the $so(2,1)$ algebra admit spin 
interpolating half-integer numbers  \cite{Bargmann:1946me,Barut:1965} and are referred to as fractional. 
The physical realisations of fractional spins are known as anyons, and their statistics interpolates 
between bosons and fermions \cite{Leinaas:1977fm,Wilczek:1982wy,Forte:1990hd}. 

As higher spin gravity aims at describing fields with arbitrary spin, for completeness, 
in three dimensions it should be extended to incorporate fundamental fractional-spin fields. 
The first step to achieve this goal was given in \cite{Boulanger:2013naa} using operator formalism. 
Later this was done in  Ref. \cite{Boulanger:2015uha} by means of deformation quantization methods 
--- i.e. using star-products \cite{Vasiliev:1999ba}. 

The model constructed in  Ref. \cite{Boulanger:2015uha} is a Chern--Simons theory 
for a gauge field that can be expressed in the form
\begin{equation}
    \label{A1} \mathbb A=\left[\begin{array}{cc} W&\psi\\[5pt] \overline\psi &U
\end{array}\right]\quad \cong \quad \left[\begin{array}{cc} \hbox{HS gravity } & \hbox{Fractional spin } \\[5pt]  
\hbox{Fractional spin } & \hbox{Internal interactions }
\end{array}\right],
\end{equation}
where the blocks correspond to four different sectors of the gauge algebra of the theory, 
$\mathcal{A}(2;\nu|w)\,$, dubbed fractional-spin algebra, which we shall introduce below. 
To these sectors we associate a higher-spin gravity connection $W\,$, an internal connection 1-form $U\,$
and the ``fractional-spin gravitinos" $(\psi, \overline\psi)\,$. 
Indeed, the Chern--Simons action obtained for \eqref{A1} resembles the Achucarro--Townsend 
theory \cite{Achucarro:1987vz} of supergravity in three dimensions. 
However, any attempt to extend standard supergravity with fractional-spin fields would lead 
to higher spin gravity. 
Since fractional-spin Lorentz representations are infinite dimensional, 
there appear infinitely many additional symmetries that can be gauged: The higher spin symmetries.
The naive matrix (super)trace gives rise, in this 
context, to divergences that cannot readily be regularised consistently without changing the 
definition of super-traces of matrices.
One achievement of \cite{Boulanger:2015uha} can be regarded as a solution to this problem; 
see \cite{Campoleoni:2013lma} and \cite{Khesin:1994ey} 
for a related discussion and an extension of the naive matrix trace.\footnote{We are
grateful to A. Campoleoni and T. Proch\'azka for discussions on this issue.}

In what follows we present the main points that lead to the formulation of fractional-spin gravity
as presented in  Ref. \cite{Boulanger:2015uha}. Here we give a somewhat simplified presentation 
compared the more technical work \cite{Boulanger:2015uha} to 
which we refer for a complete and precise treatment.  

\subsection{The fractional spins algebra}

The building block of the fractional spin algebra is the  algebra $Aq(2;\nu)$ introduced by 
Vasiliev \cite{Vasiliev:1989re} and identified with the universal enveloping algebra
of the deformed oscillator algebra \cite{Wigner:50,Yang:51} 
(see also the Refs. \cite{Plyushchay:1994re,Plyushchay:1997ty}), in turn presented by
\begin{equation}
     [q_\alpha ,q_\beta]_\star=2i(1+\nu k)\epsilon_{\alpha\beta}\ ,\qquad
\{k,q_\alpha\}_\star=0\ ,\qquad k\star k=1\ ,\label{do1}
\ee
\be
(q_\alpha)^\dagger=q_\alpha\ ,\qquad k^\dagger = k\ ,\quad \nu\in\mathbb R\ 
.\label{do2}
\end{equation}
The associative algebra $Aq(2;\nu)$ consists of arbitrary star polynomials (of finite degree) 
in $(q_\alpha,k)\,$, which in Weyl order read
\begin{equation}
    T_{\alpha(n)} := q_{\alpha_1} \cdots q_{\alpha_n}\equiv q_{(\alpha_1}\star\cdots\star q_{\alpha_n)}\ ,\qquad T_{\alpha(n)}\star k \ ,
\end{equation} 
where the symmetrisation has unit strength. 
We split the algebra in four sectors, using the projected elements
\begin{equation}
    \label{PTP}
T^{\sigma,\sigma'}_{\alpha(n)} := \left[q_{\alpha_1} \cdots q_{\alpha_n}\right]^{\sigma,\sigma'}
:=\Pi^{\sigma}\star q_{(\alpha_1}\star\cdots\star q_{\alpha_n)}\star  \Pi^{\sigma'}
\ , \qquad \Pi^\pm=\frac{1}{2} (1\pm k)\,,
\end{equation} 
which are non-vanishing iff $\sigma\sigma'=(-1)^n\,$.
These projections belong to the sub-algebras
${Aq}(2;\nu)^{\sigma,\sigma'}
=\Pi^{\sigma}\star{Aq}(2;\nu)\star\Pi^{\sigma'}\ \in \ {Aq}(2;\nu)$.
Formally, the space ${Aq}(2;\nu)$ does not contain distribution (non-polynomial) class 
of functions. It is necessary, in order to support fractional spins, 
to extend   ${Aq}(2;\nu)$ 
with certain ``$w$-class distributions", where  $w$ refers to the operator which is diagonalised by them. 
We will refer to the algebra of these elements  as $Aw(2;\nu)$. 
The operator $w$ appears in the definition of the spin operator
\be 
J_0=\frac12 \, w\star \Pi^+\ ,\qquad w:=\frac14 \, (\tau_0)^{\alpha\beta}
q_\alpha\star q_\beta\ ,\label{spinop}
\ee
which is here chosen in a non-standard form (cf. \cite{Vasiliev:1989re}), as it involves a projector $\Pi^+$.
$J_0$ generates the rotations in the spatial plane, and the projection  $\Pi^+$ plays an essential 
role in order to create fractional spins in the connection. More generally, the spin part of the Lorentz 
transformation is generated by
\be 
J_a=\frac14 \,(\tau_a)^{\alpha\beta} J_{\alpha\beta}\ ,\qquad
J_{\alpha\beta}=\frac12 \, q_{(\alpha}\star q_{\beta)}\ \star \Pi^+\  \quad \in \quad {Aq}(2;\nu)^{+,+}\,, \quad a=0,1,2 \ ,\label{Ja}
\ee
where in terms of Pauli matrices,
\be (\tau_a)^{\alpha\beta}=(\tau_a)^{\beta\alpha}=(\mathds{1},\sigma^1,\sigma^3)\ ,\qquad
(\tau^a)_{\alpha}{}^{\beta}=(\tau^a)^{\alpha'\beta}\epsilon_{\alpha'\alpha} =(-i\sigma^2,-\sigma^3,\sigma^1)\ .
\label{taumatrices}
\ee
Note that here $(\tau^a)_{\alpha}{}^{\beta}$ generates a real Clifford algebra and that the conjugation matrix $\epsilon$ 
rises and lower spinor indices. For more detail about these conventions the reader may consult the Ref. \cite{Boulanger:2015uha}.

The distribution needed to describe fractional spin are solutions of the ``star-genvalue" 
problem,
\be \label{sgen}
2J_0\star  T_E = (w\star \Pi^+ ) \star T_E = E \,T_E \;.
\ee
The operator $w$ can also be expressed in terms of ladder operators defined as
\begin{eqnarray}
\label{number} w=a^+ a^-=\tfrac{1}{2}\, \{ a^- , a^+\}_\star\ ,
\qquad
[w,a^\pm]_\star=\pm a^\pm\ ,
\end{eqnarray}
where
\be a^\pm = u^{\pm \alpha}q_\alpha\ ,\quad u^{+\alpha} u^-_\alpha~=~-\frac{i}2\ ,\quad 
(u^{\pm}_\alpha)^\dagger~=~u^\mp_\alpha\ . 
\ee
Since powers of $w$ form a closed subalgebra, as we shall see below, 
in order to solve \eqref{sgen} 
we expand their $\star$-genfunctions as\footnote{Notice that this amount to taking
$\sigma = +1$ in the conventions of Ref. \cite{Boulanger:2015uha}.}
\begin{equation}\label{choice}
T_E=\sum_{m=0}^\infty f_m w^m \star\Pi^{ +} \ , \qquad w^m=(a^+)^m  (a^-)^m \ ,
\end{equation}
where $f_m$ are constants.
We can verify that
\be
a^\pm \star \left[w^m\right]^{\sigma,\sigma}=
\left[a^\pm \left(w^m \mp \frac{m(2m+1-\nu\sigma)}{2(2m+1)} w^{m-1}\right)\right]^{-\sigma,\sigma}\ , 
\label{contraction3}
\ee
and hence
\bea
w\star \left[w^m\right]^{\sigma,\sigma}&=&\left[w^{m+1}+ \lambda^\sigma_m
w^ {m-1}\right]^{\sigma,\sigma}\ ,
\label{wcontraction}
%
\eea
where we have defined
\begin{equation}\label{lambdam}
\lambda^\sigma_m=-\frac{m^2}4 \frac{(2m+1-\nu\sigma)(2m-1+\nu\sigma)}{(2m+1)(2m-1)}
\ .
\end{equation}
%
%
Using \eqref{wcontraction} in \eqref{sgen} yields a recursive formulas. To restrict the space of solutions, 
and using our intuition on the harmonic oscillator in Fock space, we can solve the lowest 
weight condition to identify the ground state\footnote{Here we are taking the choice 
$\epsilon = +1$ in the notation of \cite{Boulanger:2015uha}.}
\be 
a^- \star T_{E_0} = 0\,,
\ee
for which we find a unique solution with $f_0=1$, given by
\begin{equation}
f_m= \frac{(-2)^m}{m!} \frac{\left(\frac32\right)_m}{\left(\frac{3-\nu}2\right)_m}\ ,
\end{equation}
where the Pochhammer symbol $(a)_n$ is given by $1$ if $n=0$ 
and by $a(a+1)\cdots (a+n-1)$ if $n=1,2,\dots$
Hence, using the definition of the confluent hypergeometric function, viz.
\be {}_1 F_1(a;b;z)=\sum_{n\geqslant 0} \frac{(a)_n}{(b)_n}\frac{z^n}{n!}\ ,\ee
we have
\begin{equation}
T_{E_0} = {}_1 F_1\left(\frac32;\frac{3-\nu}2;-2 w\right)\star \Pi^\sigma\ ,
\end{equation}
which obeys 
\be
(w-E_0)\star T_{E_0}=0\,,\qquad E_0=\frac{1+\nu}{2}\,,
\ee 
by virtue of \eqref{wcontraction}.
$T_{E_0}$ can be identified with the ground state non-normalised projector $|0 )( 0|$ of 
the \emph{deformed harmonic oscillator} introduced by Wigner \cite{Wigner:50,Yang:51}. 
Higher states can be generated by the left and right $\star$-multiplication of elements of $Aq(2;\nu)$, in terms of $\star$-powers of ladder operators,
 \be \label{kets}
 |m )( n| = (a^+)^{ \star n} \star |0 )( 0|  \star (a^-)^{\star n}\,, \qquad 
|0 )( 0|:=T_{E_0}\, ,
\ee
such that 
\be 
(w-E_m)\star |m )( n| \,=\, 0 \, =  
 |m )( n| \star (w-E_n) \,, \qquad  E_m= \big(m+\frac{1+\nu}{2}\Big)\,.
 \ee
The projector $\Pi^+$ in the definition of $J_a$ makes its action trivial on odd parity labels 
since $\Pi^+ \star |2n+1 )( m|=|m )(2n+1|\star \Pi^+ =0 \,$.

Introducing the projections of the algebra $Aw(2;\nu)$
\be Aw(2;\nu)^{\sigma,\sigma'}
=\Pi^{\sigma}\star {Aw}(2;\nu) \star\Pi^{\sigma'}\ ,
\ee
the fractional spin algebra is defined by specification of four sectors
\be\label{A2}  \mathcal{A}(2;\nu|w) := \left[\begin{array}{cc} Aq(2;\nu)^{+,+} & Aw(2;\nu)^{+,-} \\[5pt]  Aw(2;\nu)^{-,+} & Aw(2;\nu)^{-,-}
\end{array}\right] \qquad \ni  \qquad \mathbb A=\left[\begin{array}{cc} W&\psi\\[5pt] \overline\psi &U
\end{array}\right]\,,
\ee
to which the gauge connection \eqref{A1} belongs to. 

Thus, in each sector the gauge connection must be expanded as follows,
\begin{align}
\label{op}
\mathbb A = & \left[\begin{array}{cc} W&\psi\\[5pt] \overline\psi &U
\end{array}\right]\, 
\nonumber \\
= &\, \left[\begin{array}{cc} W = \sum_{n}  W^{\alpha(n)} \ 
q_{\alpha(n)}\star \Pi^+  & \psi= \sum_{m,n\geq 0}  \psi^{mn} | 2m )(
2n+1|\\[5pt] \overline\psi = \sum_{n,m\geq 0}  \overline \psi^{mn} | 2m+1 )(
2n| & U= \sum_{n,m\geq 0}  U^{mn} | 2m+1 )(
2n+1|
\end{array}\right]\,.
\end{align}
These sectors satisfy the ``fusion rules"
\bea 
\mathcal{A}(2;\nu|w)^{\sigma,\sigma'} \star \mathcal{A}(2;\nu|w)^{\tau,\tau'} \cong \delta^{\sigma',\tau} \mathcal{A}(2;\nu|w)^{\sigma,\tau'}\,,
\eea
where the projections
\be \mathcal{A}(2;\nu|w)^{\sigma,\sigma'}
=\Pi^{\sigma}\star \mathcal{A}(2;\nu|w) \star\Pi^{\sigma'}\ ,
\ee
fall in the correspondent blocks of \eqref{A2}.

Let us perform now a (global) Lorentz transformation
\be
\mathbb{A}'= g_e\star \mathbb{A}\star g_\epsilon^{-1}\, , \qquad g_\epsilon =\exp_\star (i \epsilon^a J_a )\,.
\ee
Because the projector $\Pi^+$ in its definition, $J_a$  acts non-trivially only in the upper diagonal 
block (of higher spin gravity) and on the off-diagonal blocks (``fractional spin gravitinos"), 
while the lower diagonal block $U$ (of internal interactions) does not transform. 
Internal transformations, whose basis of generators is given by $| 2m+1 )(2n+1|$ 
(up to normalisations and reality conditions), act non-trivialy on fractional spin fields and on itself.   

As we are interested to show how fractional spins make their appearance, let us perform a rotation by $2\pi\,$, 
focusing in the sector $\mathcal{A}(2;\nu|w)^{\pm , \mp}$. The parameter of transformation is given by 
$\epsilon= 2\pi J_0$, hence single elements of the basis of the sectors 
$\mathcal{A}(2;\nu |w)^{\pm , \mp}$ transform as
\bea
g_{2\pi} \star | 2m)(2n+1|\star g_{-2\pi} &=& e^{i\pi (m+\tfrac{1+\nu}{4})}  \; | 2m)(2n+1| \,, \\  
g_{2\pi} \star | 2n+1)(2m| \star g_{-2\pi}  &=& e^{-i\pi (m+\tfrac{1+\nu}{4})}  \; | 2n+1)(2m| \,,
\eea
hence it follows that the projector basis $| \hbox{odd})(\hbox{even}|\oplus |\hbox{even})(\hbox{odd}|$ 
have non-fermionic/bosonic statistical phases. Note that for odd values of $\nu$ the phases become fermionic 
or bosonic. The cases $\nu$ negative-odd are critical, in the sense that the representations of the Lorentz 
algebra become non-unitary and decouple in two sectors, of finite dimension  and non-unitary, and of infinite 
dimension and unitary. For positive-odd $\nu$ the representations of half-integer spin of the Lorentz algebra 
are unitary and infinite dimensional. Thus, for half-integer values of the spin, there appear different 
representations of the Lorentz group. Indeed, the second (infinite dimensional case) is more exotic, 
since the field theory of these type of representation is less known, while in the finite dimensional 
case the field theories are standard, including e.g. Dirac, Rarita--Schwinger, or Bargmann--Wigner, 
and well known gauge gravity models including $SL(N)$-like Chern--Simons higher spin gravities 
in three dimensions. For field theories of infinite dimensional representations with 
half-integer spins the reader can consult the  
Refs. \cite{Majorana:1932rj,Sudarshan:1970ss,Dirac:1971cy,Staunton:1974dy}. 
More details and complete analysis on the fractional spin algebra and its critical limits 
can be found in  Refs. \cite{Boulanger:2013naa,Boulanger:2015uha}.

\subsection{Gravitational and gauge couplings from Chern-Simons fractional spin gravity}

For a polynomial class of functions, with elements $f(q;k)$ Vasiliev's super-trace is given by,
\be \label{Str}
{\rm STr}_{{Aq}(2;\nu)} f(q;k)= f(0;-\nu) 
\ee
It is not straightforward that Vasiliev's super-trace \eqref{Str} will be consistent when operating 
on elements of the fractional spin algebra, because the presence of non-polynomial functions. 
It suffices to establish two consistency conditions:
(i) Finite star products, ii) Finite Vasiliev supertraces (which together with (i) implies cyclicity).

The condition (i) implies that
\be |0)(0|\star |0)(0|
 = {\cal N}^{-1}\, 
|0)(0|\ ,\qquad
{\cal N} \; \in \;\; ]0,\infty[\ ,
\label{key}\ee
where ${\cal N}$ is a normalisation constant.
This calculation was verified up to first order in $w$-power series and at all order in $\nu\,$.
Assuming that this remains true for all order in $w$ implies the existence of a normalised projector
\be 
|0\rangle \langle 0| 
:= {\cal N} \,|0)(0|\ ,\qquad |0\rangle \langle 0| \star |0\rangle \langle 0| 
=|0\rangle \langle 0|\ ,
\label{Proj0}
\ee
and the normalisation can be obtained trusting on the supertrace applied to the latter product, 
for which we need to compute just the zero order term in the $w$-expansion.
Doing so we obtain 
\be 
{\cal N}^{-1}= \frac{1 -\nu}{2}\; \;.
\ee
Before writing down the Chern--Simon action, 
we should introduce a Clifford algebra $\{ \mathds{1}$, $\Gamma\}\,$, 
$\Gamma^2=\mathds{1}\,$, which will allow us to double\footnote{The generator $\Gamma$ was 
denoted $\gamma$ in  Ref. \cite{Boulanger:2015uha}.} 
the fractional higher spin algebra 
and embed in it the $AdS_3\,$ isometry algebra $so(2,2)\cong so(2,1) \oplus  so(2,1)\,$. 
The supertrace of functions in oscillator variables and $\Gamma$ is now defined as
\be \label{Strfsg}
{\rm STr} f(q;k;\Gamma)= f(0;-\nu;0)\,. 
\ee
It can be verified that 
\be {\rm STr} \left(J_a\star J_b\right)=
\frac1{32}\,(1-\nu^2)(1-\frac\nu3)\, \eta_{ab}\ ,\ee
and
\be {\rm STr} \left(P_{m}{}^n \star
P_{m'}{}^{n'} \right) =  - \delta_m^{n'}\delta_{m'}^n\ ,\qquad P_m{}^n:= |m \rangle \langle n |
\quad  \in \quad u(\infty) \,.
\ee
%
%
%
Comparing with normalised trace operations operations, such that
\be {\rm Tr}_{\rm grav} (J_a\star J_b)=\tfrac{1}{2} \;\eta_{ab}\ ,\qquad
{\rm Tr}_{\rm int} (P_m{}^n\star P_{m'}{}^{n'})=\tfrac{1}{2} \;\delta_m^{n'}\delta_{m'}^n\ ,\label{normTr}\ee
it follows that
\begin{align} {\rm STr}|_{{\rm grav}} = \,&\tfrac{1}{16}\,(1-\nu^2)(1-\frac\nu3) \,{\rm Tr}_{\rm grav}\ ,
\\
 {\rm STr}|_{\rm int} = \, & -2 \,{\rm Tr}_{\rm int}\ ,
\end{align}
where ${\rm STr}|_{\cdot}$ means restriction of the supertrace of the fractional spin gravity 
\eqref{Strfsg} either to the gravity sector or the internal sector respectively.
Hence the Chern--Simon action for the fractional spin theory reads
\be \label{CSmodel} 
S[\mathbb A]=
\frac{\varkappa}{2\pi}\int_{M_3} {\rm Tr}_{\cal A} 
\left(\tfrac{1}{2}\, 
\mathbb A\star d\mathbb A+
\tfrac{1}{3}\, \mathbb A\star \mathbb A \star \mathbb A\right)\ ,
\ee
where the trace ${\rm Tr}_{{\cal A}}$ of 
$\mathbb{F} \in {\cal A}(2;\nu\vert w)\otimes {\rm Cliff}(\Gamma)$ 
is defined by\footnote{Out of the two possibilities for 
the fractional-spin algebra denoted by ${\cal A}_{\pm}$ in Ref. \cite{Boulanger:2015uha}, 
here we choose ${\cal A}_{+}$ that we call ${\cal A}\,$ for short. 
A proper assignment of semi-classical statistics for the components of the 
fractional-spin fields $\psi$ and $\bar \psi$
requires the introduction of a fermionic Klein operator denoted $\xi$ in 
Ref.~\cite{Boulanger:2015uha} s.t. the components of the one-form 
$\mathbb{A} \in {\cal A}(2;\nu\vert w)\otimes {\rm Cliff}(\Gamma)\otimes {\rm Cliff}(\xi)\,$ 
are all bosonic.}  
\begin{equation}
    {\rm Tr}_{\cal A}[\mathbb{F}] := {\rm Str} [ \Gamma\star ( \mathbb{F}^{+,+} + \mathbb{F}^{-,-} )]\;.
\end{equation}
With the decomposition  
\begin{align}
    W (q,k,\Gamma) = \tfrac{1+\Gamma}{2}\,\star W_L(q,k) 
    +
    \tfrac{1-\Gamma}{2}\,\star W_R(q,k) \;,\quad idem \quad U\;,
\end{align}
the action \eqref{CSmodel} produces 
\begin{align}
    S[\mathbb A] = \tfrac{1}{2}\, S[{\mathbb A}_L] - 
    \tfrac{1}{2}\, S[{\mathbb A}_R] \;,
\end{align}
where ($c=L,R$)
\begin{align}
    S[\mathbb A_{c}] = \frac{\varkappa}{2\pi}\,\int 
\left[    {\cal L}_{\rm CS}(W_{c}) +  
    {\cal L}_{\rm CS}(U_{c}) + \tfrac{1}{2}\,{\rm Str} \Big(
    \psi_{c}\star D\overline{\psi}_{c} +
    \overline{\psi}_{c}\star D{\psi}_{c}
    \Big)\right]
    \;,
\end{align}
in terms of the Chern--Simons Lagrangian
\be {\cal L}_{\rm CS}(W) =
 {\rm STr}\left[ \tfrac12 \,W \star d W+\tfrac{1}{3} \,W\star W\star W\right]\ ,
\ee
\emph{idem} ${\cal L}_{\rm CS}(U)\,$, and the covariant derivatives
\be 
D\psi=d\psi+W\star \psi+\psi\star U\ ,\qquad D\overline\psi=d\overline\psi+U\star \overline\psi
+\overline\psi\star W\ .
\ee
By comparison with the sum of the standard gravity action and gauge 
interactions in absence of fractional 
spin gravitinos and higher spin interactions, 
\begin{align}
S_{\rm grav}[W] + S_{\rm int}[U] = \;&\;
 \frac{k_{\rm grav}}{2\pi}\int_{M_3} 
{\rm Tr}_{\rm grav} \left[\tfrac12 \,W_{\rm grav}\star dW_{\rm grav} 
+\tfrac13 \, W_{\rm grav}^{\star 3}\right] 
\nonumber \\
& 
+\frac{k_{\rm int}}{2\pi}\int_{M_3}
{\rm Tr}_{\rm int} \left[\tfrac12 \, U\star dU+\tfrac13 \, U^{\star 3}\right]\;,
\end{align}
that gives, up to boundary terms in the gravitational sector, 
the sum of the Einstein--Hilbert action and an internal Chern--Simons theory for the group 
$U(\infty)\otimes U(\infty)\,$,
\begin{align} 
S_{\rm grav}[W] + S_{\rm int}[U] = \frac{k_{\rm hs}}{4\pi \ell}\int d^{3}x\,\sqrt{-g}  \left(R+\frac{2}{\ell^{2}}\right) + \frac{k_{\rm int}}{4\pi}\int_{M_3}
{\rm Tr}_{\rm int} \left( U \star dU+\tfrac{2}{3} \, U^{\star 3}\right), 
\end{align}
we find that the higher spin gravity and the internal couplings are 
given by\footnote{We take the opportunity to correct a typo appearing 
in the expression for $k_{\rm hs}$ given in the first equation (4.30) of Ref.~\cite{Boulanger:2015uha}.} 
\be
k_{\rm hs}=\frac{\varkappa}{32}\,(1-\nu^2)(1-\frac\nu 3)\ ,\qquad
 k_{\rm int}= -\varkappa \ .
 \label{couplings}
 \ee
The Newton constant is given by $G_N= \ell / (4 k_{\rm hs})\,$. 
The relation between the coupling constants of the (fractional-spin) gravity 
sector and the internal interaction sector is therefore given by
\begin{align}
 k_{\rm hs}=-\tfrac{1}{32}\,(1-\nu^2)(1-\frac\nu 3) \,k_{\rm int}\;.   
\end{align}

The interactions predicted by the model \eqref{CSmodel} can be read from the resulting equations of motion:
\be d\mathbb A+\mathbb A\star \mathbb A=0\ ,\label{eom}\ee
which in components are given by
\be dW+W\star W+\psi\star \overline\psi=0\ ,\qquad dU+U\star U+\overline \psi\star \psi=0\ ,\label{eom1}
\ee
\be d\psi+W\star \psi+\psi\star U=0\ ,\qquad d\overline\psi+U\star \overline\psi+\overline\psi\star 
W=0\ .\label{eom2Bis}
\ee
Here we observe how the fractional-spin fields source the field strength of higher-spin gravity and 
the internal interactions, while they couple minimally to the latter interactions, either from the left or 
right actions. We can visualise this result as saying that fractional-spin charges carry higher-spin gravity 
fluxes. 

To conclude this section, we would like to mention that we have omitted many details 
for the sake of simplicity. The complete treatment can be found in Ref.~\cite{Boulanger:2015uha}. 
In terms of the notation of  Ref. \cite{Boulanger:2015uha},  
here we took the choice $\sigma=+1\,$, $\epsilon=+1\,$ and considered the fractional-spin algebra 
${\cal A}_+\,$ for which the component fields in the ``gravitino'' sector are fermionic and multiplied
by the fermionic Kleinien $\xi\,$.

%
\section{Matter-coupled 3D higher-spin gravity}\label{sec:PVBlencowe}
%

In this section, we review the results of Ref.~\cite{Bonezzi:2015igv} 
where an action was provided for matter-coupled 3D higher-spin gravity, 
that reproduces upon variation the full nonlinear bosonic Prokushkin--Vasiliev (PV) equations. 
We take the opportunity to review the PV equations and  
spell out the truncation of the PV spectrum of matter fields to a single \emph{real} scalar field.
This minimal truncation  can be useful
in the context of the Gaberdiel--Gopakumar conjecture \cite{Gaberdiel:2010pz}. 
See Ref.~\cite{Gaberdiel:2012uj} for a review.

\subsection{Geometric Formulation of Prokushkin--Vasiliev's system}

In this subsection we are going to present a geometric 
formulation of Prokushkin--Vasiliev systems \cite{Prokushkin:1998bq}, 
describing matter coupled to gauge fields in three dimensional spacetime. 
Master fields consist of a one form $A$ and a zero form $B$, defined on a so-called correspondence 
space ${\cal M}_5={\cal M}_3\times {\cal Z}_2$, where ${\cal M}_3$ is the three dimensional spacetime 
manifold with local coordinates $x^\mu$, and ${\cal Z}_2$ is a non-commutative manifold with 
coordinates $z^\alpha$, $\alpha=1,2$. The fields take value in the higher spin algebra that 
extends $sp(2,\mathbb{R})$, generated by twistor variables $y^\alpha$, and are tensored with elements $\Gamma_i$ that 
generate a Clifford algebra: $\{\Gamma_i,\Gamma_j\}=2\delta_{ij}$, for $i,j=1,..,N$.
\begin{equation}
A=dx^\mu U_\mu(x,z|y;\Gamma_i)+dz^\alpha V_\alpha(x,z|y;\Gamma_i)\;,\quad B=B(x,z|y;\Gamma_i)\;.
\end{equation}
The dependence of the master fields on $(y^\alpha,z^\alpha)$ is
treated using symbol calculus, whereby they belong
to classes of functions (or distributions) on ${\cal Y}_2\times
{\cal Z}_2$ that can be composed using two associative products:
the standard commutative product rule, denoted by juxtaposition, 
and an additional noncommutative product rule, denoted by a $\star\,$.
In what follows, we shall use the normal ordered basis in which 
the star product rule is defined formally by
\begin{equation}
(f\star g)(y,z):=\int_{\mathbb{R}^4}\frac{d^2ud^2v}{(2\pi)^2}
e^{iv^\alpha u_\alpha} f(y+u,z+u)\, g(y+v,z-v)\;,
\label{hsstar}
\end{equation}
whereas a more rigorous definition requires a set of fusion 
rules.
In particular, the above composition rule rigorously
defines the associative Weyl algebra ${\rm Aq}(4)$.
This algebra consists of arbitrary polynomials in $y^\alpha$ and $z^\alpha$, modulo
\begin{equation}
y_\alpha\star y_\beta=y_\alpha y_\beta+ i\epsilon_{\alpha\beta}\ ,\qquad 
y_\alpha\star z_\beta=y_\alpha z_\beta- i\epsilon_{\alpha\beta}\ ,
\ee
\be 
z_\alpha\star y_\beta=z_\alpha y_\beta+ i\epsilon_{\alpha\beta}\ ,\qquad z_\alpha\star z_\beta
=z_\alpha z_\beta- i\epsilon_{\alpha\beta}\ ,
\ee
whose symmetric and anti-symmetric parts, respectively,
define the normal order and the (ordering independent) 
commutation rules, viz.\footnote{The doublet variables
$y^\alpha$ and $z^\alpha$ form Majorana spinors once the 
equations are cast into a manifestly Lorentz covariant form.}
\be [y^\alpha,y^\beta]_\star=-[z^\alpha,z^\beta]_\star=2i\epsilon^{\alpha\beta}\ ,\quad
 [y^\alpha,z^\beta]_\star=0\ .
 \ee
The basis one-forms $(dx^\mu,dz^\alpha)$ obey
\begin{equation}
[dx^\mu,f]_\star=0=[dz^\alpha,f]_\star\;,
\end{equation}
where the graded star commutator of differential forms is given by 
\begin{equation}
[f,g]_\star=f\star g-(-1)^{{\rm deg}(f){\rm deg}(g)}g\star f\;,
\end{equation}
with deg denoting the total form degree on ${\cal M}_3\times{\cal Z}_2\,$.
To describe bosonic models, we impose 
\begin{equation} \pi(A)=A\;,\quad \pi(B)=B\; \end{equation}
where $\pi$ is the automorphism of the differential star product algebra defined by
\begin{equation}
\pi(x^\mu,dx^\mu,z^\alpha, dz^\alpha,y^\alpha,\Gamma_i)
=(x^\mu,dx^\mu,-z^\alpha, -dz^\alpha,-y^\alpha,\Gamma_i)\;.
\end{equation}
The hermitian conjugation is defined by 
\begin{equation}
(f\star g)^\dagger = 
(-1)^{\rm{deg}(f)\rm{deg}(g)} g^\dagger \star f^\dagger\ ,\qquad 
(z_\alpha,dz^\alpha;y_\alpha,\Gamma_i)^\dagger=(-z_\alpha,-dz^\alpha;y_\alpha,\Gamma_i)\ .
\end{equation}
and the reality conditions on the master fields read
\be
A^\dagger = -A\ ,\qquad B^\dagger = B\ .
\ee
Defining
\be F= dA+ A\star A\ ,\quad D B= dB+ A\star B-B\star A\ ,\quad
d=dx^\mu\partial_\mu+dz^\alpha\frac{\partial}{\partial z^\alpha}\ ,
\ee
where the differential obeys
\begin{eqnarray}
d( f\star g)~=~(d f)\star g+(-1)^{\rm deg(f)} 
f\star d g\ ,
\quad 
(df)^\dagger=d(f^\dagger)\;,
\end{eqnarray}  
the PV field equations can be written as
\begin{equation}\label{geomPV}
\begin{split}
&F+ B\star J=0\ ,\qquad DB=0\;,
\end{split}
\end{equation}
where 
\be 
J:=-\tfrac{i}{4}\,dz^\alpha dz_\alpha\,\kappa\;\qquad 
\kappa:=e^{iy^\alpha z_\alpha}\;.
\ee
The element $J$ is closed and central in the space of $\pi$-invariant forms, viz.
\be dJ=0\ ,\qquad J\star f=\pi(f)\star J\ ,\ee
as can be seen from the fact that $\kappa$, 
which is referred to as the inner Klein operator, 
obeys
\be \kappa\star f(x,dx,z,dz,y,\Gamma_i)\star\kappa=f(x,dx,-z,dz,-y,
\Gamma_i)\ .
\end{equation}
It follows that \eqref{geomPV} defines a universally Cartan 
integrable system (i.e. a set of generalized curvature
constraints compatible with $d^2\equiv 0$ in any dimension).
The Cartan gauge transformations take the form
\begin{equation}
\delta_\epsilon A= d\epsilon+[A,\epsilon]_\star\;,\quad\delta_\epsilon B=[B,\epsilon]_\star \;.
\end{equation}
In order to see the equivalence with Prokushkin--Vasiliev systems, let us introduce the oscillator-like fields $S_\alpha:=z_\alpha-2i\,V_\alpha$ and split the field equations in $dx$ and $dz$ directions, thus obtaining
\begin{equation}\label{oscPV}
\begin{split}
&d_XU+U\star U=0\;,\quad d_XB+[U,B]_\star=0\;,\quad d_XS_\alpha+[U,S_\alpha]_\star=0\;,\\[2mm]
&[S_\alpha,B]_\star=0\;,\quad[S_\alpha,S_\beta]_\star=-2i\epsilon_{\alpha\beta}\,\big(1-B\star\kappa\big)\;,
\end{split}
\end{equation}
where $d_X:=dx^\mu\partial_\mu$ is the spacetime differential.
We stress that, due to the bosonic projection, one has $\{S_\alpha,\kappa\}_\star=0$, that will be crucial for the discussion of massive vacua.

\paragraph{Massless vacua.}

Let us analyse the above system around the vacuum solution $B_0=0\,$. From the vacuum equation $[S_{0\alpha}, S_{0\beta}]_\star=-2i\epsilon_{\alpha\beta}$ one can take  $S_{0\alpha}=z_\alpha$ and hence
\begin{equation}
0=[z_\alpha,U_0]_\star=-2i\,\frac{\partial U_0}{\partial z_\alpha}\quad\rightarrow\quad U_0=\Omega(x|y;\Gamma_i)\;,
\end{equation}
and the remaining equation is the flatness condition $d\Omega+\Omega\star\Omega=0\,$.
Bilinears in $y$ variables generate $sp(2,\mathbb{R})$ under star-commutators:
\begin{equation}
T_{\alpha\beta}:=\frac{1}{2i}y_\alpha y_\beta\;,\quad
[T_{\alpha\beta},T_{\gamma\delta}]_\star=4\epsilon_{(\alpha(\gamma}\,T_{\beta)\delta)}\;,
\end{equation}
but some outer element is needed in order to double $sp(2,\mathbb{R})$ and thus represent the $AdS_3$ isometry algebra $sp(2,\mathbb{R})\oplus sp(2,\mathbb{R})$. 
In order to describe massless vacua, it turns out that the minimal dimension of the Clifford algebra is $N=2\,$, 
and one can write the vacuum spacetime connection as
\begin{equation}
\label{ads3}
\Omega(x|y;\Gamma_i)=\frac{1}{4i}\Big(\omega^{\alpha\beta}(x)\,y_\alpha y_\beta 
+\Gamma_1\,e^{\alpha\beta}(x)\,y_\alpha y_\beta\Big)\equiv \omega+\Gamma_1\,e\;,
\end{equation}
where $\omega^{\alpha\beta}$ and $e^{\alpha\beta}$ are the background Lorentz connection and dreibein, 
respectively. The flatness condition amounts then to
\begin{equation}
de_{\alpha\beta}+2\omega_{(\alpha}{}^\gamma\wedge e_{\beta)\gamma}=0\;,\quad d\omega_{\alpha\beta}+\omega_\alpha{}^\gamma\wedge \omega_{\beta\gamma}=-e_\alpha{}^\gamma\wedge e_{\beta\gamma}\;,
\end{equation}
that indeed describes $AdS_3$ spacetime with
unit radius. 
In order to study fluctuations around this vacuum we expand the master fields as $U=\Omega+w_1+..$, 
$B=0+B_1+..$ and $S_\alpha=z_\alpha+S_{1\alpha}+..$, and the field equations for first order fluctuations 
read
\begin{equation}
\begin{split}
&D_0w_1=0\;,\quad D_0B_1=0\;,\quad D_0S_\alpha+2i\,\frac{\partial w_1}{\partial z^\alpha}=0\;,\\
&\frac{\partial B_1}{\partial z^\alpha}=0\;,\quad \frac{\partial S_1^\alpha}{\partial z^\alpha}=B_1\star\kappa\;,
\end{split}
\end{equation}
from which we see that $B_1=C(x|y;\Gamma_i)$ is $z$-independent, 
and the background covariant derivative is defined by 
$D_0f:=df+[\Omega,f]_\star$, with $\Omega$ given by \eqref{ads3}. 
If we make explicit the dependence on $\Gamma_i$ in $C$ as
\begin{equation}
C(x|y;\Gamma_i)=C_{\rm aux}(x|y;\Gamma_1)+C_{\rm dyn}(x|y;\Gamma_1)\,\Gamma_2\;,
\end{equation}
we can see that $D_0C=0$ splits into
\begin{equation}
D_LC_{\rm aux}+\Gamma_1[e,C_{\rm aux}]_\star=0\;,\quad 
D_LC_{\rm dyn}+\Gamma_1\{e,C_{\rm dyn}\}_\star=0\;,
\end{equation}
where the Lorentz covariant derivative is 
$D_L:=d+[\omega,\bullet]_\star\,$. 
It is known \cite{Vasiliev:1992gr} that the equation for 
$C_{\rm aux}$ describes non-propagating degrees of freedom, and $C_{\rm aux}$ is 
indeed referred to as auxiliary. On the other hand, by expanding 
the equation for $C_{\rm dyn}$ in power series in $y^\alpha\,$, one can see that 
$C_{\rm dyn}$ contains two real AdS$_3$ massless scalar fields 
together with all their on-shell nontrivial 
derivatives. At this stage, the next steps would be to
\begin{itemize}
\item[(1)] use $\frac{\partial S_1^\alpha}{\partial z^\alpha}=C\star\kappa$ 
to solve $S_{1\alpha}$ in terms of $C\,$, up to some pure-gauge contribution; 
\item[(2)] substitute for $S_1^\alpha$ into $\frac{\partial w_1}{\partial z^\alpha}=-D_0S_{1\alpha}$ 
to solve the $z$-dependence of $w_1$ in terms of $C\,$. 
One gets $w_1=\omega_1(x|y;\Gamma_i)+f_1(x,z|y;\Gamma_i)$, where $f_1$ is a known function linear in $C\,$;
\item[(3)] use $D_0\omega_1=-\Big(D_0f_1\Big)\rvert_{z=0}$ as higher spin equation for $\omega_1\,$. 
\end{itemize}
It is known \cite{Vasiliev:1992ix}, however, that the linearized curvature $D_0\omega_1$ does not receive 
nontrivial sources linear in $C$, and indeed the higher spin fields do not propagate. 
For a more complete and thorough analysis of the 3D equations, including the issue of field redefinition, 
locality and elimination of the auxiliary zero-form $C_{\rm aux}$ up to second order in the weak fields, 
see Ref.~\cite{Kessel:2015kna}. See also Ref.~\cite{Iazeolla:2015tca} for another review 
and exact solutions of the PV equations. 

\paragraph{Massive vacua.}

Let us study the model around a different vacuum, namely $B_0=\nu\,\Gamma$, where $\nu$ 
is a constant 
and $\Gamma$ is some element generated by the $\Gamma_i\,$, 
obeying $\Gamma^\dagger=\Gamma$ and $\Gamma^2=1$. The vacuum equations now read
\begin{equation}
\begin{split}
&dU_0+U_0\star U_0=0\;,\quad [\Gamma, U_0]=0\;,\quad[\tilde z_\alpha,\Gamma]_\star=0\\
&[\tilde z_\alpha,U_0]_\star=0\;,\quad [\tilde z_\alpha,
\tilde z_\beta]_\star=-2i\,\epsilon_{\alpha\beta}\big(1-\nu\,\Gamma\kappa\big)\;,
\end{split}
\end{equation}
where we defined $\tilde z_\alpha:=S_{0\alpha}$, that obey\footnote{Since $\tilde z_\alpha$ commutes with 
$\Gamma$ and anticommutes with $\kappa$ from the necessary bosonic projection.} 
$\{\tilde z_\alpha,\Gamma\kappa\}_\star=0$ and the  deformed oscillator algebra. 
The strategy is first to find a solution for $\tilde z_\alpha$, and then solve $[\tilde z_\alpha,U_0]_\star=0$ 
by finding other deformed variables $\tilde y_\alpha$ that star commute with $\tilde z_\alpha$. 
To this end, let us introduce, as done in Ref.~\cite{Prokushkin:1998bq},
\begin{equation}
\sigma_\alpha:=\nu\int_0^1dt\,t\,e^{ityz}(y_\alpha+z_\alpha)\;,\quad \tau_\alpha:=\nu\int_0^1dt\,(t-1)\,e^{ityz}(y_\alpha+z_\alpha)\;,
\end{equation}
with $yz:=y^\alpha z_\alpha$. One can check that they obey the following relations:
\begin{equation}
\begin{split}
&[z_{[\alpha},\sigma_{\beta]}]_\star=-i\nu\,\epsilon_{\alpha\beta}\,\kappa\;,\quad [\sigma_\alpha,\sigma_\beta]_\star=0\;,\quad \{\sigma_\alpha,\tau_\beta\}_\star=0\;,\\
&[z_\alpha,\tau_\beta]_\star=\{\sigma_\alpha,y_\beta\}_\star\;,\quad \{y_{[\alpha},\tau_{\beta]}\}_\star=i\nu\,\epsilon_{\alpha\beta}\;,\quad [\tau_\alpha,\tau_\beta]_\star=0\;.
\end{split}
\end{equation}
Let us make the following Ansatz for the deformed oscillators:
\begin{equation}
\tilde z_\alpha=X\,z_\alpha+Y\,\sigma_\alpha\;,\quad \tilde y_\alpha=A\,y_\alpha+B\,\tau_\alpha\;,
\end{equation}
where $X$, $Y$, $A$ and $B$ are built out of gamma matrices $\Gamma_i$. 
By demanding 
\begin{equation}
[\tilde z_\alpha,\tilde z_\beta]_\star=-2i\epsilon_{\alpha\beta}(1-\nu\Gamma\kappa)~, 
\quad [\tilde y_\alpha,\tilde z_\beta]_\star=0~,
\end{equation}
one has the following constraints, 
\begin{equation}
X^2=1\;,\quad XY=YX=-\Gamma\;,\quad [A,X]=\{A,Y\}=[B,X]=\{B,Y\}=0\;.
\end{equation}
As in the original Prokushkin--Vasiliev model, we demand that $\tilde y_\alpha$ obey a deformed oscillator algebra:
\begin{equation}
[\tilde y_\alpha,\tilde y_\beta]_\star=2i\epsilon_{\alpha\beta}\big(1-\nu\,\Gamma\big)\;,\quad \{\tilde y_\alpha,\Gamma\}=0\;.
\end{equation}
This imposes
\begin{equation}
\{A,\Gamma\} = 0\;,\quad  A^2 = 1\;,  \quad AB = -BA = -\Gamma\;.   
\end{equation}
%
A convenient solution, that admits propagating degrees of freedom, is given by $\Gamma=\Gamma_{1234}$
$$
X=1\;,\quad Y=-\Gamma\;,\quad A=\Gamma_1\;,\quad B=-\Gamma_{234}=-\Gamma_1\Gamma\;,
$$
where we have chosen the number of generators of the Clifford algebra to be $N=4\,$ and $\Gamma_{i_1..i_k}:=\Gamma_{[i_1}..\Gamma_{i_k]}$.

With this solution, fields obeying $[f,\tilde z_\alpha]_\star=0$ and $[f,\Gamma]=0$ are 
given by $f(x|\tilde y;\Gamma_{ij})$, i.e. star functions of the deformed $\tilde y$'s and of all 
bilinears $\Gamma_{ij}$, together with $\Gamma$ itself, that is generated by bilinears.
One can check that undeformed $sp(2,\mathbb{R})$ is still generated by $\tilde y$'s 
as $T_{\alpha\beta}:= \frac{1}{4i}\{\tilde y_\alpha,\tilde y_\beta\}_\star$, and one can take the background 
connection $U_0=\Omega$ to be
\begin{equation}\label{defads3}
\Omega(x|\tilde y;\Gamma_{ij})=\frac{1}{4i}\Big(\omega^{\alpha\beta}(x)\,\tilde y_\alpha\star\tilde y_\beta+i\Gamma_{23}\,e^{\alpha\beta}(x)\,\tilde y_\alpha\star\tilde y_\beta\Big)=\omega+i\Gamma_{23}\,e\;,
\end{equation}
such that $d\Omega+\Omega^{\star 2}=0$ is solved by the $AdS_3$ background.
Fluctuations $B_1$ defined by $B=\nu\,\Gamma+B_1+..$ obey the linearized equations
\begin{equation}
[\tilde z_\alpha,B_1]_\star=0\quad\Rightarrow \quad B_1=C(x|\tilde y;\Gamma_{ij})\;,\quad D_0C=0\;,
\end{equation}
where, as before, $D_0f:=df+[\Omega,f]_\star$, but now with $\Omega$ given by \eqref{defads3}. 
In order to find the propagating degrees of freedom, let us explicitate the $\Gamma_{ij}$ dependence 
of $C$:
\begin{equation}
C(x|\tilde y;\Gamma_{ij})=C_{\rm aux}(x|\tilde y;\Gamma,i\Gamma_{23})+C_{\rm dyn}(x|\tilde y;\Gamma,i\Gamma_{23})i\Gamma_{24}\;,
\end{equation}
and use it in the equation $D_0C=0$, obtaining
\begin{equation}
D_LC_{\rm aux}+i\Gamma_{23}\,[e,C_{\rm aux}]_\star=0\;,\quad  D_LC_{\rm dyn}
+i\Gamma_{23}\,\{e,C_{\rm dyn}\}_\star=0\;,
\end{equation}
from which one can find that $C_{\rm dyn}$ contains four real massive propagating scalars, 
while $C_{\rm aux}$ yields four topological deformations. 
The analysis is now equivalent to the original Prokushkin--Vasiliev model, 
since at this level one can identify 
\begin{equation} 
\left(k\right)_{\rm PV}=\Gamma\;,\quad
\left(\nu\right)_{\rm PV}=-\nu\;,\quad
\left(\rho\right)_{\rm PV}=\Gamma_1\;,\quad 
\left(y_\alpha\right)_{\rm PV}=\Gamma_1\,y_\alpha\;,\quad \left(z_\alpha\right)_{\rm PV}=\Gamma_1\,z_\alpha\;,\label{map1}
\end{equation}
\be \left(\psi_1\right)_{\rm PV}=i\Gamma_{23}\ ,\qquad \left(\psi_2\right)_{\rm PV}=i\Gamma_{24}\ .
\label{map2}\ee
One can truncate the master fields as
\begin{equation}
\Pi_\Gamma^+A=A\;,\quad \Pi_\Gamma^+B=B\;,
\end{equation}
with the projector $\Pi_\Gamma^+:=\frac{1\pm\Gamma}{2}\;,$ 
yielding a model with two propagating scalars contained in 
\begin{equation}
C_{\rm dyn}(x|\tilde y;\Gamma,i\Gamma_{23})=\frac{1+\Gamma}{2}\,C_+(x|\tilde y;i\Gamma_{23})\ .
\end{equation}

\paragraph{Minimal truncation.}

We now want to further truncate the matter sector contained in $C_+$ to a single, real scalar field.
By using the anti-automorphism defined by 
\be \tau(f\star g) = (-1)^{\rm{deg}(f)\rm{deg}(g)} \tau(g) \star \tau(f)\ ,\ee
\be 
\tau(z^\alpha,dz^\alpha;y^\alpha,\Gamma_i)
=(-iz^\alpha,-idz^\alpha;iy^\alpha,\epsilon_{(i)} \Gamma_i)\ ,\qquad \epsilon_{(i)}=(+,+,-,-)\ ,
\ee
one can truncate further, by requiring
\begin{equation}
\tau(A)=-A\;,\quad \tau(B)=B\;.    
\end{equation}
This last truncation leaves a single real propagating 
scalar in the spectrum, since now the two chains of fields 
multiplying $1$ and $i\Gamma_{23}$ only contain even and 
odd numbers of derivatives of the physical field:
\begin{equation}
C_{+}(x|\tilde y;i\Gamma_{23})=\sum_{n=0}^\infty C^0_{\alpha_1...\alpha_{4n}}(x)\,\tilde  y^{\alpha_1}...\tilde y^{\alpha_{4n}} +i\Gamma_{23}\sum_{n=0}^\infty C^1_{\alpha_1...\alpha_{4n+2}}(x)\,\tilde  y^{\alpha_1}...\tilde y^{\alpha_{4n+2}}\;.
\end{equation}

Finally, it is possible to find solution for the deformed 
oscillator algebra such that the deformed 
oscillators obey the same reality conditions and 
transformation properties under the tau map as the 
undeformed $y^\alpha$ and~$z^\alpha\,$, and therefore it 
is understood above that the master fields are expanded 
in terms of these deformed oscillators.

\subsection{Covariant Hamiltonian action}
\label{sec:BulkAction}
%

In this section we begin by discussing some generalities on covariant Hamiltonian actions on ${\cal X}_4\times {\cal Z}_2$. We then determine the constraints on the Hamiltonian such that it leads to a master action in which the master field content, including the Lagrange multipliers, are extended to consist of sum of even and odd forms of appropriate degree, and central elements. This action yields a generalized version of the PV field equations.

\paragraph{Generalities.} 

In order to formulate the theory 
within the AKSZ framework \cite{Alexandrov:1995kv} using its
adaptation to noncommutative higher spin geometries proposed
in \cite{Boulanger:2012bj}, we assume a formulation of the PV 
system that treats ${\cal Z}_2$ as being closed and introduce 
an open six-manifold ${\cal M}_6$ with boundary
\be \partial{\cal M}_6={\cal X}_3\times{\cal Z}_2\ ,\ee
where ${\cal X}_3$ is a closed manifold containing ${\cal M}_3$ 
as an open submanifold. 
On ${\cal M}_6$, we introduce a two-fold duality extended
\cite{Vasiliev:2007yc,Boulanger:2011dd,Vasiliev:2015mka}
\footnote{Starting from a universally 
Cartan integrable system and replacing each $p$-form by a sum of forms 
of degrees $p$, $p+2$, $\dots$, $p+2N$, and each structure constant 
by a function of off-shell closed and central terms,
i.e. elements in the de Rham cohomology valued in 
the center of the fiber algebra, with a decomposition
into degrees $0$, $2$, $\dots$, $2N$, yields a new universally Cartan 
integrable system, referred to as the $N$-fold
duality extension of the original system.
More generally, one may consider on-shell duality
extensions by including on-shell closed complex-valued 
functionals into the extension of the structure constants 
\cite{Sezgin:2012ag,Vasiliev:2015mka}.} set of differential forms given by
\begin{eqnarray}
A &=& A_{[1]}+A_{[3]}+A_{[5]}\;,\qquad 
B = B_{[0]}+B_{[2]}+B_{[4]}\;,
\\
T &=& T_{[4]}+T_{[2]}+T_{[0]}\;,\qquad S = S_{[5]}+S_{[3]}+S_{[1]}
\;,
\end{eqnarray}
valued in ${\cal A}\otimes {\cal C}^+_4$, ${\cal C}^+_4$ being the even subspace of the four dimensional Clifford algebra,  and where the subscript denotes the form degree. The restriction to the even Clifford subalgebra, i.e. to fields obeying $[f(\Gamma_i),\Gamma]=0$, is required by demanding integrability of the field equations coming from the action that we will present in the following.
We let $\{ J^{I}\}$ denote the generators of the ring 
of off-shell closed and central terms, i.e.
elements in the de Rham cohomology of ${\cal M}_6$
valued in the center of ${\cal A}\otimes {\cal C}^+_4$, which hence obey
\begin{eqnarray}
{d} J^I~=~0\ ,\quad \left[ J^I, f\right]_\star~=~0\ ,
\end{eqnarray} 
(off-shell) for any differential form $f$ on ${\cal M}_6$
valued in ${\cal A}\otimes {\cal C}^+_4$.
Following the approach of \cite{Boulanger:2011dd}, we consider actions
of the form
\begin{eqnarray}
S_{\rm H}~&=&~\int_{{\cal M}_6} {\rm Tr}_{{\cal A}\otimes {\cal C}^+_4}
\left[  S\star D B+ T\star  F+{\cal V}( S, T; B; J^I)\right]
\label{action1}
\\
&=&\int_{{\cal M}_6}{\rm Tr}_{{\cal A}\otimes {\cal C}^+_4}
\left[  S\star {d} B+ T\star  {d}A-{\cal H}( S, T;A, B; J^I)\right]
\label{action2}
\end{eqnarray}
where ${\rm Tr}_{{\cal A}\otimes {\cal C}^+_4}$ denotes
a cyclic trace operation 
on ${\cal A}\otimes {\cal C}^+_4\,$.
We assume a structure group gauged by $A$ and that 
$S$, $T$ and $B$ belong to sections, and \eqref{action2} 
makes explicit the covariant Hamiltonian form, with
\begin{equation}
\begin{split}
&{\cal H}(S,T;A,B;J^I)=-S\star[A,B]_\star-T\star A\star A-{\cal V}(S,T;B;J^I)\;.
\end{split}
\end{equation}
Thus, the coordinate and momentum master fields, defined by
\be  
(X^\alpha;P_\alpha):=(A,B;T,S)\ ,
\ee
lie in subspaces of ${\cal A}$ that are dually paired 
using ${\rm Tr}_{\cal A}$, which leads to distinct 
models depending on whether these subspaces are isomorphic 
or not. 
In the reductions that follow, we shall consider the
first type of models, moreover, for definiteness, we shall assume that
\be 
{\cal M}_6={\cal X}_4\times {\cal Z}_2\ ,\ee
and the associative bundle is chosen such that
\be 
\check {\cal L}=\oint_{{\cal Z}_2} {\rm Tr}_{{\cal A}\otimes {\cal C}^+_4}
\left[  S\star D B+ T\star  F+{\cal V}( S, T; B; J^I)\right]\ ,
\ee
is finite and globally defined on ${\cal X}_4$.
The action can then be written as
\be S_{\rm H} = \int_{{\cal X}_4} 
\check{\cal L}\ .
\ee
%

\paragraph{The master action.}

The Hamiltonian is constrained by gauge invariance, 
or equivalently, by universal on-shell Cartan integrability\footnote{
Covariant Hamiltonian actions are gauge invariant iff their 
equations of motion form universally Cartan integrable systems.}.
In addition, it is constrained by the requirement
that the equations of motion on ${\cal M}_6$ 
reduce to a desired set of equations of motion 
on $\partial{\cal M}_6$ upon assuming 
natural boundary conditions.
In order to obtain a model that admits consistent truncations
to three-dimensional CS higher spin gravities, we need to 
assume that ${\cal V}$ contains a term that is quadratic in $T$.
The simplest possible such action is given by
\be
S_{\rm H} = \int_{{\cal M}_6} {\rm Tr}_{{\cal A}\otimes {\cal C}^+_4}
\left[ S\star DB + T\star\left[F+g+h\star ( B - \tfrac12\mu\star T) \right] 
+\mu \star B\star S\star S \right]
\label{ma}
\ee
where 
\be 
g=g(J^I)\ ,\qquad h=h(J^I)\ ,\qquad \mu=\mu(J^I)
\ee
are even closed and central elements on ${\cal M}_6$
in degrees
\be {\rm deg}(g,h,\mu)=(2\ \mbox{mod $2$},2\ \mbox{mod $2$},0\ \mbox{mod $2$})\ .\ee
The reality conditions are given by
\be (A,B;T,S;g,h,\mu)^\dagger=(-A,B;-T,S;-g,-h,-\mu)\ ,\ee
The total variation yields
\begin{eqnarray}
\delta S_{\rm H} &=& \int_{{\cal M}_6} 
{\rm Tr}_{{\cal A}\otimes{\cal C}^+_4}\Big( 
\delta T \star{\cal R}^A +\delta S\star {\cal R}^B +\delta A\star {\cal R}^T +\delta B \star{\cal R}^S\Big)
\nonumber\\
&& +\;\oint_{\partial{\cal M}_6}{\rm Tr}_{{\cal A}\otimes{\cal C}^+_4}(T\star\delta A - S\star \delta B) \;,
\end{eqnarray}
where the Cartan curvatures read

\medskip

\be
\begin{aligned}
{\cal R}^A &=F + g+ h\star ( B - \mu\star T)\approx 0 \\
{\cal R}^B &= DB + \mu \star [S,B]_\star\approx 0 \\
{\cal R}^T &= DT + [S,B]_\star\approx 0 \\
{\cal R}^S &= DS +h\star T+ \mu \star S\star S\approx0
\label{e4}
\end{aligned}
\ee
\medskip
The generalized Bianchi identities are
\begin{eqnarray}
D{\cal R}^A &\equiv & h\star ( {\cal R}^B  - \mu\star {\cal R}^T)  \;,
\\
D{\cal R}^B &\equiv & [ ( {\cal R}^A + \mu\star{\cal R}^S ) , B]_\star - \mu\star\{ {\cal R}^B , S \}_\star\;,
\\ 
D{\cal R}^T &\equiv & [ {\cal R}^A , T]_\star + [  {\cal R}^S , B]_\star  - \{ {\cal R}^B , S \}_\star\;,
\\ 
D{\cal R}^S &\equiv & [  {\cal R}^A , S ]_\star + \mu\star [ {\cal R}^S , S ]_\star + h\star{\cal R}^T \;.
\end{eqnarray}
The gauge transformations 
\begin{eqnarray}
\delta_{\epsilon,\eta} A &=& D\epsilon^A - h\star(\epsilon^B -\mu\star \eta^T)\;,
\\
\delta_{\epsilon,\eta} B &=& D\epsilon^B - [\epsilon^A , B]_\star - \mu\star[\eta^S , B]_\star + \mu \star
\{ S,\epsilon^B\}_\star \;,
\\
\delta_{\epsilon,\eta} T &=& D\eta^T - [\epsilon^A , T]_\star - [\eta^S , B]_\star+ \{ S,\epsilon^B\}_\star  \;,
\\
\delta_{\epsilon,\eta} S &=& D\eta^S - [\epsilon^A , S]_\star - \mu\star [\eta^S , S]_\star -  h\star \eta^T\;,
\end{eqnarray}
which transform the Cartan curvatures into each other, induce 
\begin{eqnarray}
\delta_{\epsilon,\eta}S_{\rm H} &=& \oint_{\partial{\cal M}_6} {\rm Tr}_{{\cal A}\otimes {\cal C}^+_4}
\Big( \eta^T\star [F+ g+h\star B] + \eta^S \star DB \Big)\;.
\end{eqnarray}
We take $(\epsilon^B;\eta^T,\eta^S)$ to belong 
to sections of the structure group and impose\footnote{
Following the AKSZ approach, the Batalin--Vilkovisky classical
master equation requires that the ghosts corresponding
to $(\eta^T,\eta^S)$ vanish at $\partial{\cal M}_6$ off-shell.}
\be (\eta^T,\eta^S)\rvert_{\partial{\cal M}_6}=0\ .\ee
We have also assumed that $(A,B)$ fluctuate on $\partial{\cal M}_6$, which implies
\be T\rvert_{\partial{\cal M}_6}\approx 0\approx S\rvert_{\partial{\cal M}_6}\ .\label{bc}\ee
The resulting boundary equations of motion 
\begin{equation}
F+g+ h\star B\approx0\;,\quad DB\approx0\;
\label{ev}
\end{equation}
thus provide a duality extended version of the 
Prokushkin--Vasiliev equations, that is free from any interaction 
ambiguity, following a variational principle.
Let us notice that in the action \eqref{ma}, the relative coefficient of the
$BSS$ and $TT$ terms is fixed uniquely by Cartan integrability.
%

\subsection{Consistent truncations}\label{Sec:Blencowe}

In this section we review \cite{Bonezzi:2015igv} the consistent truncation of the above 
covariant Hamiltonian master action in six dimensions down
to a $BF$-like model on ${\cal X}_4$ that reproduces Blencowe's action.
The truncation consists in integrating out the fluctuations in 
$B$ around its vacuum expectation value
followed by reductions on ${\cal X}_4\,$.
Starting from the equations of motion \eqref{e4} and setting
$B=0$ yields 
\be F + g- h\star \mu\star T=0 \ ,\qquad  DT = 0\ ,\label{sub}\ee
and
\be 
DS +h\star T+ \mu \star S\star S=0\ , 
\label{ds}
\ee
which together form a Cartan integrable system containing 
\eqref{sub} as a subsystem, i.e.
the free differential algebra generated by $(A,T,S)$
contains a subalgebra generated by $(A,T)$.
Assuming $\partial {\cal M}_6$ to consist 
of a single component, it follows from 
$S\vert_{\partial {\cal M}_6} =0$ that $S$
can be reconstructed from $(A,T)$ on-shell
\footnote{
Since $T\vert_{\partial {\cal M}_6} =0$
on-shell as well it follows that both $S$
and $T$ can be taken to vanish on ${\cal M}_6$
on-shell.} from \eqref{ds}.
Therefore, the system \eqref{sub} is a consistent 
truncation of the original system \eqref{e4} on-shell.

Rewriting the full action \eqref{ma} by integrating
by parts in its $SDB$-term yields
\be 
S_{\rm H}=\int_{{\cal M}_6} {\rm Tr}_{{\cal A}\otimes {\cal C}^+_4}
\left[T\star (F+g-\tfrac12 h\star \mu \star T)+B\star (DS+h\star T+\mu\star S\star S)\right]\ .
\ee
It follows that $B=0$ is a saddle point of
the path integral at which $B$ and $S$
can be integrated out in a perturbative
expansion.
Schematically, modulo gauge fixing, one has
\be \int_{\langle B\rangle =0} [DB][DS] e^{\tfrac{i}\hbar S_{\rm H}}
\sim  e^{\tfrac{i}\hbar S_{\rm eff}[A,T]}\ ,\ee
where the effective action
\be S_{\rm eff}[A,T]=S_{\rm red}[A,T]+ O(\hbar)\ ,\label{Seff}\ee
consists of loop corrections (comprising 
attendant functional determinants on noncommutative 
manifolds) and 
\be S_{\rm red}=\int_{{\cal M}_6} {\rm Tr}_{{\cal A}\otimes {\cal C}^+_4}
T\star (F+g-\tfrac12 h\star \mu \star T)\ .\label{Sred1}\ee
The latter is a consistently reduced classical action
in the sense that it reproduces the subsystem \eqref{sub}.
The reduced system, which thus consists of the 
gauge sector of the original system, is a topological 
theory with local symmetries
\begin{equation}
\delta A=D\epsilon+\mu\star h\star\eta\;,\quad\delta T
=D\eta-[\epsilon,T]_\star\;,
\end{equation}
and equations of motion and boundary conditions given by
\bea
&& F+g-\mu\star h\star T=0\;,\quad DT=0\;,
\\
&& T\rvert_{\partial{\cal M}_6}=0\;.
\label{tbc}
\eea
The boundary equations are thus given by 
\be (F+g)\rvert_{\partial{\cal M}_6}=0\ .\ee

To address Blencowe's theory, we truncate once 
more by reducing \eqref{sub} under the assumptions 
that   
\be 
g=0\ ,\quad \mu =\mu_{[0]}\equiv \mu_0\ ,\quad
h=J\ ,
\label{reduceg}
\ee
where $\mu_0$ is an imaginary constant, and
\be \label{r1}
A = \check{W}_{[1]} - \check K_{[1]}-\mu_0 J \star \check{K}_{[1]}\; , \quad  
T = \check{T}_{[2]}+ \check K_{[1]}\star \check K_{[1]} -
\mu_0 J\star \check{T}_{[2]}\;,
\ee
where we define checked fields to be $z$-independent:
$
\check{f}\in \Omega({\cal X}_4)\otimes{\cal A}\otimes 
{\cal C}^+_4 \; .
$
By defining 
\be \check F=d_X \check{W}+\check{W}\star \check{W}\ ,\quad 
\check D\check K=d_X \check{K}+[\check{W},\check{K}]_\star\ ,\quad 
\check D\check T=d_X \check{T}+[\check{W},\check{T}]_\star\ ,\label{defcheck}\ee
and suppressing the subscripts indicating form degrees,
the reduction of \eqref{sub} yields 
\be 
\check F+ \check  T=0\ ,\qquad 
\check D\check  T=0\ ,\label{red1}
\ee
\be 
D \check K-\check K\star \check K+\check T=0\ ,
\label{red2}
\ee
which is a Cartan integrable system containing \eqref{red1}
as a subsystem.
From \eqref{tbc} and \eqref{r1}, we deduce the boundary conditions 
\be \check  T\rvert_{\partial{\cal X}_4}=0=(\check K\star \check K)\rvert_{\partial{\cal X}_4}\ ,
\label{red3}\ee
Substituting \eqref{r1} into \eqref{Sred1} 
and using \eqref{red3} we obtain 
\be
\check S_{\rm red}[\check W,\check T]=-\mu_0 \int_{{\cal X}_4} \int_{{\cal Z}_2}
{\rm Tr}_{{\cal A}\otimes {\cal C}^+_4}
J\star \check {T}\left( \check{F}+
\tfrac12 \check{T}\right)\ .\label{checkSred}\ee
At this stage, we truncate the models further as follows: 
\be
\check W=\Pi^+_\Gamma W_+ + \Pi^-_\Gamma W_-\; ,\quad 
\check T=\Pi^+_\Gamma T_+ + \Pi^-_\Gamma T_- \;,
\label{Blred2}
\ee
where $W_\pm$ and $T_\pm$ are independent of $\Gamma_i\,$.
Inserting \eqref{Blred2} into \eqref{checkSred} and using 
\be 
\int_{{\cal Z}_2}  {\rm Tr}_{{\cal A}\otimes{\cal C}^+_4} J\star \Pi^\pm _{\Gamma}\check{f}
=: \pm \tfrac{i\pi}{2} \, {\rm STr}_{{\rm Aq}(2)}\,\check f\ ,
\ee
for $\check f$ independent of $\Gamma_i$ and belonging to ${\rm Aq}(2)\,$,  
the associative algebra given by the universal enveloping algebra of the undeformed 
oscillators $y^\alpha\,$. 
We thereby obtain the following four-dimensional Hamiltonian extension
of Blencowe's action:
\be
\label{BlenH}
S_{\rm Bl}= - \tfrac{i\pi}{2} \, \mu_0 \int_{{\cal X}_4} 
{\rm STr}_{{\rm Aq}(2)} \left[
 T_+( F_+ +
\tfrac12 T_+ )- T_-( F_- +
\tfrac12  T_- )\right]\; .
\ee
Assuming that ${\cal X}_4={\cal X}_3\times [0,\infty[$ and that
all fields fall off at ${\cal X}_3\times\infty$, and assuming
furthermore that ${\cal X}_3$ has a simple topology, the elimination of the Lagrange multipliers yields
\be
\label{BlenCS}
S_{\rm Bl}=\tfrac{i\pi}2\, \mu_0  \left(S_{\rm CS}[ W_+]-
S_{\rm CS}[ W_-]\right)\ ,\ee
with 
\be S_{\rm CS}[W]=\oint_{{\cal X}_3}{\rm STr}_{{\rm Aq}(2)} \Big[\tfrac12 W\star 
dW+\tfrac13W\star W\star W\Big]\;,
\ee
where now $d$ denotes the exterior derivative on ${\cal X}_3$.
Equivalently, 
\begin{equation}\label{EHaction}
S_{\rm Bl}=i\mu_0 \pi\oint_{{\cal X}_3} {\rm STr}_{{\rm Aq}(2)} \Big[E
\star (d\Omega+\Omega\star \Omega)+\tfrac13\,E\star E\star E\Big]\;,\qquad W_\pm =\Omega\pm E\ ,
\end{equation}
from which we identify
\begin{equation}
\mu_0 = -\frac{4i}{\pi^2} \frac{\ell_{\rm AdS}}{G_{\rm N}} 
\end{equation}
using the conventions of \cite{Boulanger:2015uha}. 

\paragraph{Truncation to the Prokushkin--Segal--Vasiliev action.}

In the paper \cite{Bonezzi:2015igv} that we are reviewing here, another consistent truncation of the action 
\eqref{ma} is shown to reproduce the action principle \cite{Prokushkin:1999gc} 
for the PV equation that is formulated 
in a 2-dimensional base space for the $z^\alpha$ oscillators. The interested reader can find all the 
details in \cite{Bonezzi:2015igv}.

\section{Higher spins and topological strings}\label{sec:TOS}

In this section we shall argue, based on the formal 
structure of higher spin equations and actions, that 
higher spin dynamics can be described in terms of 
first-quantized topological open strings.

Vasiliev's equations exhibit two basic properties akin to 
open string field theory:
First of all, we recall that in Cartan's formulation of field theories, 
as free differential algebras, the 
spaces of forms in degrees one and zero correspond to
 gauge Lie algebras and spectra of local degrees of 
freedom, respectively.
In this respect, a remarkable feature of Vasiliev's higher 
spin gravities is that their one- and zero-form modules 
are unified into associative algebras\footnote{
An interesting consequence is that the spectra of massless 
particles and generalized Type D solutions in 
four-dimensional higher spin gravity are related 
by a $\mathbb Z_2$-symmetry \cite{Iazeolla:2011cb}, 
unlike in ordinary gauge theories, in which the Type 
D solution spaces are finite dimensional.}.
Second, they can be embedded into a Frobenius--Chern--Simons 
theory with a cubic action principle that only refers
to the trace and star product operations of the associative 
algebra and a nilpotent differential containing
the de Rham differential.
These algebraic structures can be naturally 
encoded into a class of two-dimensional 
Poisson sigma models with gauged supersymmetry,
corresponding to the de Rham operator,
which are our candidate topological open
string models.

In order to explain the rational behind these 
models in more detail, let us start from
a particle on a symplectic manifold $N$
with symplectic potential one-form 
$\vartheta=d\phi^\alpha \vartheta_{\alpha}$
and Hamiltonian $H$, consisting of constraints
and Lagrange multipliers, as described
by an action of the form
$$ S_{\vartheta} = \int_{C} (\vartheta - H)\ ,$$
integrated along a worldline $C$. 
Assigning physical states of the system 
to Hilbert spaces it is natural to consider 
open worldlines whose boundaries are
allowed to fluctuate in Lagrangian submanifolds 
of $N$.
Alternatively, and more generally, one may
assign the physical states to density matrices, 
in which case it is natural to take $C$ to be closed,
and choose boundary conditions such that the
path integral provides a trace operation.
Considering initially the case of a trivial Hamiltonian,
and letting $\Pi^{\alpha\beta}$ denote the Poisson bi-vector
on $N$, \emph{i.e.} the inverse of the symplectic
two-form $\omega_{\alpha\beta}=\partial_\alpha \vartheta_\beta-\partial_\beta \vartheta_\alpha$,
the resulting path integral over worldlines $C$
can be extended into a path integral over
open worldsheets $\Sigma$ with boundary $C$,
weighted by $\exp \tfrac i\hbar S_{\Pi}$ where $S_\Pi$
is the action of the 
Ikeda--Schaller--Strobl Poisson sigma model
$$ S_{\Pi}=\int_{\Sigma} (\eta_\alpha \wedge d\phi^\alpha 
+\tfrac{1}{2}\, \Pi^{\alpha\beta} \eta_\alpha \wedge \eta_\beta )\ ,$$
subject to the boundary condition $\eta_\alpha|_C=0$.
(which can be derived on-shell using the variational
principle but that actually must be imposed off-shell as well
in order for the classical Batalin--Vilkovisky master
equation to hold).
The resulting path integral can be performed in two steps:
First over discs with a point of $C$ attached to a fixed point 
$p_0\in N$ (which one may think of as a $(D-1)$-brane),
and then by integrating over $p_0$ using the symplectic measure.
Inserting vertex operators along points $p_i\in C$, given by 
(the pull-back of) functions $f_i$ on $N$, the resulting path integral 
can be viewed as a formal definition of ${\rm Tr}_N \prod^\star_i f_i$,
where $ f\star g$ is an associative noncommutative product,
referred to as the star product.
Provided that $f$ and $g$ belong to sufficiently smooth classes
of functions (\emph{e.g.} polynomials or formal power series
with coefficients given by power series in $\hbar$), 
then the star product has an expansion in terms of
pointwise derivatives, \emph{viz.}
$$f\star g=fg + i\hbar \{f,g\}+\cdots\ ,\qquad \{f,g\}:= \Pi^{\alpha\beta}
\partial_\alpha f \partial_\beta g\ ,$$
where the higher order terms in the $\hbar$ expansion 
are given in terms of multiple
derivatives of the functions and the Poisson structure (but not
its inverse),
while in order to compose more general elements, including
(nonperturbative) distributions on $N$ and (which arise as 
density matrices related to unitary representations), 
nonlocal versions of the star product, based
on auxiliary integrations, are required.
Unlike the particle action, the Poisson sigma model remains 
well-defined on Poisson manifolds, where $\Pi^{\alpha\beta}$ is assumed to be a
bi-vector obeying 
$$\Pi^{\delta[\alpha}\partial_\delta \Pi^{\beta\gamma]}=0\ ,$$
as this ensures the invariance of the action under 
$\eta_\alpha$ gauge transformations, while there is
no requirement on its invertibility.

Historically, the existence of a star product on
general symplectic manifolds was first established
formally by De Wilde and Lecomte \cite{DeWildeLecomte}.
Its pointwise form was given explicitly and on a manifestly 
covariant form by Fedosov \cite{Fedosov:1994zz}.
His construction 
resembles that of Vasiliev, though it does not
provide any dynamics, which instead will require
a suitable gauging leading to topological
open strings, as we shall propose below.
Later, Kontsevich used his formality theorem
to show the existence and uniqueness (up to
similarity transformations and changes of basis
corresponding to different ordering schemes) of
the star product on general Poisson manifolds.
He also provided an explicit formula in the case of 
$N\cong \mathbb{R}^n$, derived soon after by means of perturbative AKSZ quantization of the Poisson sigma model by Cattaneo and Felder \cite{Cattaneo:1999fm},
who also provided a globally defined star product 
for any $(N,\Pi)$ in \cite{Cattaneo:2001vu}. 

In order to spell out our proposal, we need one
more basic ingredient of Vasiliev's models 
and their FCS generalizations.
In addition to the aforementioned fusion of their Cartan modules
into associative fiber algebras, their simplicity 
relies on yet one more unification, namely of spacetime and  
noncommutative symplectic manifolds into 
extended base manifolds of Poisson type so as to facilitate the construction of 
closed and central two-forms whose star products
with the zero-form provide (nontrivial)
co-cycles for the curvatures. 
The resulting framework is that of differential Poisson 
manifolds, which are natural generalizations of Poisson 
manifolds on which classes of differential forms (and distributions) 
in different degrees, and not just functions, can be equipped by 
star products by deforming the classical wedge product
along differential Poisson brackets.
In the case of sufficiently smooth objects, the resulting
pointwise star product reads
\be f\star g=f\wedge g + i\hbar \{f,g\}+\cdots\ ,\ee
where the differential Poisson bracket, that now involves the bi-vector
as well as a compatible connection, is of the form\footnote{Strictly
speaking, there remains one possible tensorial deformation;
see \cite{Arias:2015wha} for a more detailed form of the Poisson bracket. } 
\be 
\{ \omega, \eta \} = 
\Pi^{\alpha \beta}\, \nabla_\alpha \omega \wedge \nabla_\beta \eta
+ (-1)^{\deg(\omega)} \, \widetilde R^{\alpha\beta} \wedge  i_\alpha \omega \wedge i_\beta \eta \ ,\ee
where $\widetilde R^{\alpha\beta}=\Pi^{\alpha\gamma}\widetilde R^{\alpha}{}_{\gamma}$ and 
$\widetilde R^{\alpha}{}_{\beta}$ is the curvature of the connection shifted 
by the torsion.
When written on the above form, the bracket is graded skew-symmetric, of vanishing intrinsic degree,
compatible with the de Rham differential and obeying Leibniz rule, while the graded Jacobi identity requires the additional conditions
\footnote{The quadratic constraint on the curvature tensor
is a Yang--Baxter equation and the two-dimensional 
differential Poisson sigma model can thus be used
as a framework for associative algebras including
Hopf algebras.}
\be
\Pi^{\delta [\alpha}T^\beta_{\delta\epsilon} \Pi^{\gamma]\epsilon} =  0   ~,  \qquad 
\Pi^{\alpha \rho} \Pi^{\sigma \beta} R_{\rho \sigma}{}^{\gamma}{}_\delta= 0 ~, \ee\be
\Pi^{\alpha \lambda} \nabla_\lambda \widetilde R_{\beta \gamma}{}^{\rho \sigma} = 0 ~, \qquad
\widetilde R_{\epsilon [ \rho}{}^{(\alpha \beta} \widetilde R_{\sigma \lambda]}{}^{\gamma) \epsilon} = 0 ~.
\ee
By extending the formality theorem to the graded 
supermanifold with coordinates
$(\phi^\alpha,\theta^\alpha)$ (see \cite{McCurdy:2009xz}
for a related discussion), the $\hbar$-expansion of the star product can be determined
together with a nilpotent operator, given by an
$\hbar$-deformation of the de Rham differential,
by requiring associativity and compatibility.
In particular, in degree zero, the star product must yield Cattaneo and Felder's  manifestly covariant form of Kontsevich product (which in its turn reproduces Fedosov's product in the symplectic case).

Motivated by the above considerations, it was proposed in \cite{Arias:2015wha}
that the differential star product algebra can be obtained by perturbative quantization of the supersymmetric two-dimensional topological sigma model
based on the (classical) action
\be
S=\int_{\Sigma}
\left(  
\eta_\alpha \wedge d\phi^\alpha 
+ \chi_\alpha \wedge \nabla \theta^\alpha 
+\tfrac{1}{2}\, \Pi^{\alpha\beta} \eta_\alpha \wedge \eta_\beta 
+\tfrac{1}{4} \theta^\alpha \theta^\beta \widetilde R_{\alpha\beta}{}^{\gamma\delta} \,\chi_\gamma \wedge \chi_\delta 
 \right)\ .
\ee
Indeed, the conditions on the background fields
required by the Jacobi identity, as listed above, 
can equivalently be derived by requiring the appropriate 
two-dimensional gauge symmetries, which
serve to gauge away all local degrees of freedom except 
the constant modes in $\phi^\alpha$ and $\theta^\alpha$.
Moreover, the coefficient of the 
four-fermi coupling is fixed by the requirement that
the action has a 
global nilpotent supersymmetry given by 
\be \delta_{\rm f} \phi^\alpha=\theta^\alpha\ ,\qquad \delta_{\rm f} \theta^\alpha=0 ,\ee
\be \delta_{\rm f} \eta_\alpha =\tfrac{1}{2} \widetilde R_{\beta\gamma}{}^{\delta}{}_{\alpha} \, 
\chi_\delta \,\theta^\beta 
\theta^\gamma - \Gamma^\gamma_{\alpha\beta} \, \eta_\gamma \, \theta^\beta  \ ,\qquad
\delta_{\rm f}\chi_\alpha = -\eta_\alpha + \Gamma^{\gamma}_{\alpha\beta} \, \chi_\gamma \, 
\theta^\beta\ .
\ee
This transformations can be identified as an avatar for the de 
Rham differential upon representing the forms 
on $N$ as functions on the parity shifted bundle 
$T[1]N$ coordinatized by $(\phi^\alpha,\theta^\alpha)$.
It can be shown that the model
can be reformulated on more general $(n|n)$ 
supermanifolds equipped with super Poisson
bi-vectors.
The global symmetries of this model, which include the 
original Hamiltonian vector fields as well as more 
general super Killing vectors, such as $\delta_{\rm f}$,
can be gauged.

The special case of gauging of $\delta_{\rm f}$ was
studied in \cite{Bonezzi:2015lfa}.
It remains to be examined whether it is 
consistent at the quantum level, which may 
require extra conditions on differential Poisson 
geometry (\emph{e.g.} its Ricci tensor), 
though the absence of obstructions
at the first sub-leading order in $\hbar$ \cite{McCurdy:2009xz} 
suggests their absence to all orders in perturbation theory.
Assuming quantum consistency, it is natural to propose that
finite deformations of the background can be modelled by a 
topological open string field $\Psi$ obeying
\begin{equation} 
\{ Q, \Psi\}_\star + \Psi\star \Psi=0\ ,
\end{equation}
where the BRST operator $Q$ contains a 
sector that gauges the de Rham differential
acting on the zero-modes of $\phi^\alpha$.
We claim that the FCS formulation of four-dimensional
higher spin gravity in twistor space can be obtained by a reduction 
of the above system down to a finite set of modes
describing stretched strings, whereas in order to
obtain the minimal bosonic models\footnote{These models
are known as the $D$-dimensional Type A models.
The $D$-dimensional Type B models should arise in
a similar fashion from gauging an $osp(1|2)$ algebra.} 
based on vector oscillators 
one has to gauge the additional $sp(2)$ Killing vectors 
generated by the moment functions of the conformal particle.

Finally, let us remark on the connection to tensionless strings in
anti-de Sitter spacetime.
To this end, it is important that the
associative algebras in higher spin gravity are of a special 
type related to singletons.
These algebras can be realized either via the group 
algebras (or enveloping algebras) of the underlying 
finite-dimensional isometry groups or oscillator 
algebras over ideals given by singleton annihilators
\footnote{The topological open string approach may be 
useful in further elucidating the whether there exist
two dual underlying first-quantized formulations, one on 
the group manifold, thought of as a Poisson manifold, 
and another one directly on the singleton
phase space, thought of as a symplectic manifold.}, 
which can be thought of as the equations
of motions for conformal particles on the embedding space
of anti-de Sitter spacetime.
Thus one may view the topological open string as the
germ of a tensionless string, consisting of two string
partons orbiting (as conformal particles) around a 
center-of-mass.
Indeed, similar excitations, carrying the quantum numbers
of singletons, are known to arise in the form of cusps on 
tensionful closed boconic strings in $D$-dimensional 
anti-de Sitter space time in the semi-classical limit 
(described by soliton solutions of the of the Nambu--Goto
action). 
Thus, as proposed in \cite{Engquist:2005yt}, one
may think of the Hagedorn transition in flat spacetime
as switching on a small negative cosmological
constant whereby the cyclically ordered one-string 
states of tensionful strings break up in the tensionless 
limit into totally symmetric multi-singleton states.
In particular, on physical grounds, long (folded) 
string states, which thus connect two cusps at the opposite
side of the center-of-mass (with closed worldlines in periodic
anti-de Sitter spacetime), should remain self-interacting in this limit, 
at least in the classical limit of the string field theory, and 
admit a first-quantized description as an $sp(2)$-gauged topological open 
bosonic string with en effective 
field theory description in terms of minimal bosonic higher spin gravity.

In summary, one further piece of motivation for studying topological open
strings of the type proposed above is thus that they may
provide a link between higher spin gravity and tensionless
strings in anti-de Sitter spacetime, and possibly also a glimpse
of what one might expect from a background independent
formulation of closed string field theory. 
%

\section*{Acknowledgements}

We thank Th. Basile, A. Campoleoni, V. Didenko, 
C. Iazeolla, E. Skvortsov, Ph. Spindel, T. Proch\'azka 
and M. Vasiliev for discussions. 
N.B., Associate F.R.S.-FNRS Researcher, 
is very grateful to the Institute of Advanced Studies from Nanyang Technological University
in Singapore for hospitality, and to the organisers of the workshop ``Higher Spin Gauge Theories'' for
providing the opportunity to present this work and contribute to the proceedings. 
The work of E.S. is supported in part by NSF grant PHY-1214344.
P. S. would like to acknowledge the support of
CONICYT grant DPI 20140115 and FONDECYT grant Regular
1140296.

\providecommand{\href}[2]{#2}\begingroup\raggedright\endgroup


\begin{thebibliography}{10}

\bibitem{Arias:2015wha}
C.~Arias, N.~Boulanger, P.~Sundell, and A.~Torres-Gomez, ``{2D sigma models and
  differential Poisson algebras},''
  \href{http://dx.doi.org/10.1007/JHEP08(2015)095}{{\em JHEP} {\bf 08} (2015)
  095},
\href{http://arxiv.org/abs/1503.05625}{{\tt arXiv:1503.05625 [hep-th]}}.

\bibitem{Boulanger:2015uha}
N.~Boulanger, P.~Sundell, and M.~Valenzuela, ``{Gravitational and gauge
  couplings in Chern-Simons fractional spin gravity},''
  \href{http://dx.doi.org/10.1007/JHEP01(2016)173}{{\em JHEP} {\bf 01} (2016)
  173},
\href{http://arxiv.org/abs/1504.04286}{{\tt arXiv:1504.04286 [hep-th]}}.

\bibitem{Boulanger:2015kfa}
N.~Boulanger, E.~Sezgin, and P.~Sundell, ``{4D Higher Spin Gravity with
  Dynamical Two-Form as a Frobenius--Chern--Simons Gauge Theory},''
\href{http://arxiv.org/abs/1505.04957}{{\tt arXiv:1505.04957 [hep-th]}}.

\bibitem{Bonezzi:2015igv}
R.~Bonezzi, N.~Boulanger, E.~Sezgin, and P.~Sundell, ``{An Action for Matter
  Coupled Higher Spin Gravity in Three Dimensions},''
\href{http://arxiv.org/abs/1512.02209}{{\tt arXiv:1512.02209 [hep-th]}}.

\bibitem{Sezgin:2011hq}
E.~Sezgin and P.~Sundell, ``{Geometry and Observables in Vasiliev's Higher Spin
  Gravity},'' \href{http://dx.doi.org/10.1007/JHEP07(2012)121}{{\em JHEP} {\bf
  07} (2012)  121},
\href{http://arxiv.org/abs/1103.2360}{{\tt arXiv:1103.2360 [hep-th]}}.

\bibitem{Boulanger:2011dd}
N.~Boulanger and P.~Sundell, ``{An action principle for Vasiliev's
  four-dimensional higher-spin gravity},''
  \href{http://dx.doi.org/10.1088/1751-8113/44/49/495402}{{\em J.Phys.} {\bf
  A44} (2011)  495402},
\href{http://arxiv.org/abs/1102.2219}{{\tt arXiv:1102.2219 [hep-th]}}.

\bibitem{Achucarro:1987vz}
A.~Achucarro and P.~Townsend, ``{A Chern-Simons Action for Three-Dimensional
  anti-De Sitter Supergravity Theories},''
\href{http://dx.doi.org/10.1016/0370-2693(86)90140-1}{{\em Phys.Lett.} {\bf
  B180} (1986)  89}.

\bibitem{Prokushkin:1998bq}
S.~F. Prokushkin and M.~A. Vasiliev, ``{Higher-spin gauge interactions for
  massive matter fields in 3D AdS space-time},''
  \href{http://dx.doi.org/10.1016/S0550-3213(98)00839-6}{{\em Nucl. Phys.} {\bf
  B545} (1999)  385},
\href{http://arxiv.org/abs/hep-th/9806236}{{\tt arXiv:hep-th/9806236}}.

\bibitem{Blencowe:1988gj}
M.~P. Blencowe, ``{A Consistent Interacting Massless Higher Spin Field Theory
  in $D$ = (2+1)},''
\href{http://dx.doi.org/10.1088/0264-9381/6/4/005}{{\em Class. Quant. Grav.}
  {\bf 6} (1989)  443}.

\bibitem{Campoleoni:2010zq}
A.~Campoleoni, S.~Fredenhagen, S.~Pfenninger, and S.~Theisen, ``{Asymptotic
  symmetries of three-dimensional gravity coupled to higher-spin fields},''
  \href{http://dx.doi.org/10.1007/JHEP11(2010)007}{{\em JHEP} {\bf 1011} (2010)
   007}, \href{http://arxiv.org/abs/1008.4744}{{\tt arXiv:1008.4744 [hep-th]}}.

\bibitem{Gaberdiel:2010pz}
M.~R. Gaberdiel and R.~Gopakumar, ``{An $AdS_3$ Dual for Minimal Model CFTs},''
  \href{http://dx.doi.org/10.1103/PhysRevD.83.066007}{{\em Phys.Rev.} {\bf D83}
  (2011)  066007}, \href{http://arxiv.org/abs/1011.2986}{{\tt arXiv:1011.2986
  [hep-th]}}.

\bibitem{Gaberdiel:2012uj}
M.~R. Gaberdiel and R.~Gopakumar, ``{Minimal Model Holography},''
  \href{http://dx.doi.org/10.1088/1751-8113/46/21/214002}{{\em J.Phys.} {\bf
  A46} (2013)  214002},
\href{http://arxiv.org/abs/1207.6697}{{\tt arXiv:1207.6697 [hep-th]}}.

\bibitem{Bonezzi:2015lfa}
R.~Bonezzi, P.~Sundell, and A.~Torres-Gomez, ``{2D Poisson Sigma Models with
  Gauged Vectorial Supersymmetry},''
  \href{http://dx.doi.org/10.1007/JHEP08(2015)047}{{\em JHEP} {\bf 08} (2015)
  047},
\href{http://arxiv.org/abs/1505.04959}{{\tt arXiv:1505.04959 [hep-th]}}.

\bibitem{Cattaneo:2001bp}
A.~S. Cattaneo and G.~Felder, ``{Poisson sigma models and deformation
  quantization},'' \\
  \href{http://dx.doi.org/10.1142/S0217732301003255}{{\em Mod.
  Phys. Lett.} {\bf A16} (2001)  179--190},
  \href{http://arxiv.org/abs/hep-th/0102208}{{\tt arXiv:hep-th/0102208}}.

\bibitem{Vasiliev:1990en}
M.~A. Vasiliev, ``{Consistent equation for interacting gauge fields of all
  spins in (3+1)-dimensions},''
\href{http://dx.doi.org/10.1016/0370-2693(90)91400-6}{{\em Phys. Lett.} {\bf
  B243} (1990)  378--382}.

\bibitem{Vasiliev:1992av}
M.~A. Vasiliev, ``{More on equations of motion for interacting massless fields
  of all spins in (3+1)-dimensions},''
\href{http://dx.doi.org/10.1016/0370-2693(92)91457-K}{{\em Phys. Lett.} {\bf
  B285} (1992)  225--234}.

\bibitem{Vasiliev:2003ev}
M.~A. Vasiliev, ``{Nonlinear equations for symmetric massless higher spin
  fields in (A)dS(d)},''
  \href{http://dx.doi.org/10.1016/S0370-2693(03)00872-4}{{\em Phys. Lett.} {\bf
  B567} (2003)  139--151},
\href{http://arxiv.org/abs/hep-th/0304049}{{\tt arXiv:hep-th/0304049}}.

\bibitem{Fronsdal:1978rb}
C.~Fronsdal, ``{Massless Fields with Integer Spin},''
\href{http://dx.doi.org/10.1103/PhysRevD.18.3624}{{\em Phys. Rev.} {\bf D18}
  (1978)  3624}.

\bibitem{Vasiliev:1986td}
M.~A. Vasiliev, ``{Free massless fields of arbitrary spin in the de Sitter
  space and initial data for a higher spin superalgebra},''
{\em Fortsch.Phys.} {\bf 35} (1987)  741--770.

\bibitem{Lopatin:1987hz}
V.~Lopatin and M.~A. Vasiliev, ``Free massless bosonic fields of arbitrary spin
  in d-dimensional de sitter space,''
  \href{http://dx.doi.org/10.1142/S0217732388000313}{{\em Mod.Phys.Lett.} {\bf
  A3} (1988)  257}.

\bibitem{Vasiliev:2001wa}
M.~A. Vasiliev, ``{Cubic interactions of bosonic higher spin gauge fields in
  AdS(5)},'' \href{http://dx.doi.org/10.1016/S0550-3213(01)00433-3}{{\em Nucl.
  Phys.} {\bf B616} (2001)  106--162},
\href{http://arxiv.org/abs/hep-th/0106200}{{\tt arXiv:hep-th/0106200}}.

\bibitem{Fradkin:1986qy}
E.~S. Fradkin and M.~A. Vasiliev, ``{Cubic Interaction in Extended Theories of
  Massless Higher Spin Fields},''
\href{http://dx.doi.org/10.1016/0550-3213(87)90469-X}{{\em Nucl. Phys.} {\bf
  B291} (1987)  141}.

\bibitem{Vasilev:2011xf}
M.~Vasiliev, ``{Cubic Vertices for Symmetric Higher-Spin Gauge Fields in
  $(A)dS_d$},'' \href{http://dx.doi.org/10.1016/j.nuclphysb.2012.04.012}{{\em
  Nucl.Phys.} {\bf B862} (2012)  341--408},
\href{http://arxiv.org/abs/1108.5921}{{\tt arXiv:1108.5921 [hep-th]}}.

\bibitem{Buchbinder:2006eq}
I.~L. Buchbinder, A.~Fotopoulos, A.~C. Petkou, and M.~Tsulaia, ``{Constructing
  the cubic interaction vertex of higher spin gauge fields},''
  \href{http://dx.doi.org/10.1103/PhysRevD.74.105018}{{\em Phys. Rev.} {\bf
  D74} (2006)  105018},
\href{http://arxiv.org/abs/hep-th/0609082}{{\tt arXiv:hep-th/0609082}}.

\bibitem{Metsaev:2006ui}
R.~R. Metsaev, ``{Gravitational and higher-derivative interactions of massive
  spin 5/2 field in (A)dS space},''
  \href{http://dx.doi.org/10.1103/PhysRevD.77.025032}{{\em Phys. Rev.} {\bf
  D77} (2008)  025032},
\href{http://arxiv.org/abs/hep-th/0612279}{{\tt arXiv:hep-th/0612279}}.

\bibitem{Fotopoulos:2007yq}
A.~Fotopoulos, N.~Irges, A.~C. Petkou, and M.~Tsulaia, ``{Higher-Spin Gauge
  Fields Interacting with Scalars: The Lagrangian Cubic Vertex},''
  \href{http://dx.doi.org/10.1088/1126-6708/2007/10/021}{{\em JHEP} {\bf 10}
  (2007)  021},
\href{http://arxiv.org/abs/0708.1399}{{\tt arXiv:0708.1399 [hep-th]}}.

\bibitem{Zinoviev:2008ck}
Y.~M. Zinoviev, ``{On spin 3 interacting with gravity},''
  \href{http://dx.doi.org/10.1088/0264-9381/26/3/035022}{{\em Class. Quant.
  Grav.} {\bf 26} (2009)  035022},
\href{http://arxiv.org/abs/0805.2226}{{\tt arXiv:0805.2226 [hep-th]}}.

\bibitem{Boulanger:2008tg}
N.~Boulanger, S.~Leclercq, and P.~Sundell, ``{On The Uniqueness of Minimal
  Coupling in Higher-Spin Gauge Theory},''
  \href{http://dx.doi.org/10.1088/1126-6708/2008/08/056}{{\em JHEP} {\bf 0808}
  (2008)  056},
\href{http://arxiv.org/abs/0805.2764}{{\tt arXiv:0805.2764 [hep-th]}}.

\bibitem{Boulanger:2011qt}
N.~Boulanger, E.~Skvortsov, and Y.~Zinoviev, ``{Gravitational cubic
  interactions for a simple mixed-symmetry gauge field in AdS and flat
  backgrounds},'' \href{http://dx.doi.org/10.1088/1751-8113/44/41/415403}{{\em
  J.Phys.A} {\bf A44} (2011)  415403},
  \href{http://arxiv.org/abs/1107.1872}{{\tt arXiv:1107.1872 [hep-th]}}.

\bibitem{Joung:2012hz}
E.~Joung, L.~Lopez, and M.~Taronna, ``{Generating functions of
  (partially-)massless higher-spin cubic interactions},''
  \href{http://dx.doi.org/10.1007/JHEP01(2013)168}{{\em JHEP} {\bf 1301} (2013)
   168},
\href{http://arxiv.org/abs/1211.5912}{{\tt arXiv:1211.5912 [hep-th]}}.

\bibitem{Boulanger:2012dx}
N.~Boulanger, D.~Ponomarev, and E.~Skvortsov, ``{Non-abelian cubic vertices for
  higher-spin fields in anti-de Sitter space},''
  \href{http://dx.doi.org/10.1007/JHEP05(2013)008}{{\em JHEP} {\bf 1305} (2013)
   008},
\href{http://arxiv.org/abs/1211.6979}{{\tt arXiv:1211.6979 [hep-th]}}.

\bibitem{Joung:2013nma}
E.~Joung and M.~Taronna, ``{Cubic-interaction-induced deformations of
  higher-spin symmetries},''
  \href{http://dx.doi.org/10.1007/JHEP03(2014)103}{{\em JHEP} {\bf 1403} (2014)
   103},
\href{http://arxiv.org/abs/1311.0242}{{\tt arXiv:1311.0242 [hep-th]}}.

\bibitem{Bekaert:2010hw}
X.~Bekaert, N.~Boulanger, and P.~A. Sundell,
  \href{http://dx.doi.org/10.1103/RevModPhys.84.987}{``How higher-spin gravity
  surpasses the spin-two barrier,''{\em Rev. Mod. Phys.} {\bf 84} (Jul, 2012)
  987--1009},
\href{http://arxiv.org/abs/1007.0435}{{\tt arXiv:1007.0435 [hep-th]}}.

\bibitem{Alexandrov:1995kv}
M.~Alexandrov, M.~Kontsevich, A.~Schwartz, and O.~Zaboronsky, ``{The Geometry
  of the master equation and topological quantum field theory},''
  \href{http://dx.doi.org/10.1142/S0217751X97001031}{{\em Int. J. Mod. Phys.}
  {\bf A12} (1997)  1405--1430},
\href{http://arxiv.org/abs/hep-th/9502010}{{\tt arXiv:hep-th/9502010}}.

\bibitem{Boulanger:2012bj}
N.~Boulanger, N.~Colombo, and P.~Sundell, ``{A minimal BV action for Vasiliev's
  four-dimensional higher spin gravity},''
  \href{http://dx.doi.org/10.1007/JHEP10(2012)043}{{\em JHEP} {\bf 1210} (2012)
   043},
\href{http://arxiv.org/abs/1205.3339}{{\tt arXiv:1205.3339 [hep-th]}}.

\bibitem{Vasiliev:1988sa}
M.~A. Vasiliev, ``{Consistent equations for interacting massless fields of all
  spins in the first order in curvatures},''
\href{http://dx.doi.org/10.1016/0003-4916(89)90261-3}{{\em Annals Phys.} {\bf
  190} (1989)  59--106}.

\bibitem{Doroud:2011xs}
N.~Doroud and L.~Smolin, ``{An Action for higher spin gauge theory in four
  dimensions},'' \href{http://arxiv.org/abs/1102.3297}{{\tt arXiv:1102.3297
  [hep-th]}}.

\bibitem{Colombo:2010fu}
N.~Colombo and P.~Sundell, ``{Twistor space observables and quasi-amplitudes in
  4D higher spin gravity},''
  \href{http://dx.doi.org/10.1007/JHEP11(2011)042}{{\em JHEP} {\bf 1111} (2011)
   042},
\href{http://arxiv.org/abs/1012.0813}{{\tt arXiv:1012.0813 [hep-th]}}.

\bibitem{Colombo:2012jx}
N.~Colombo and P.~Sundell, ``{Higher Spin Gravity Amplitudes From Zero-form
  Charges},''
\href{http://arxiv.org/abs/1208.3880}{{\tt arXiv:1208.3880 [hep-th]}}.

\bibitem{Didenko:2012tv}
V.~Didenko and E.~Skvortsov, ``{Exact higher-spin symmetry in CFT: all
  correlators in unbroken Vasiliev theory},''
  \href{http://dx.doi.org/10.1007/JHEP04(2013)158}{{\em JHEP} {\bf 1304} (2013)
   158},
\href{http://arxiv.org/abs/1210.7963}{{\tt arXiv:1210.7963 [hep-th]}}.

\bibitem{Engquist:2005yt}
J.~Engquist and P.~Sundell, ``{Brane partons and singleton strings},''
  \href{http://dx.doi.org/10.1016/j.nuclphysb.2006.06.040}{{\em Nucl. Phys.}
  {\bf B752} (2006)  206--279},
\href{http://arxiv.org/abs/hep-th/0508124}{{\tt arXiv:hep-th/0508124}}.

\bibitem{Bekaert:2015tva}
X.~Bekaert, J.~Erdmenger, D.~Ponomarev, and C.~Sleight, ``{Quartic AdS
  Interactions in Higher-Spin Gravity from Conformal Field Theory},''
\href{http://arxiv.org/abs/1508.04292}{{\tt arXiv:1508.04292 [hep-th]}}.

\bibitem{Chu:1997ik}
C.-S. Chu and P.-M. Ho, ``{Poisson algebra of differential forms},''
  \href{http://dx.doi.org/10.1142/S0217751X97002929}{{\em Int.J.Mod.Phys.} {\bf
  12} (1997)  5573--5587},
\href{http://arxiv.org/abs/q-alg/9612031}{{\tt arXiv:q-alg/9612031 [q-alg]}}.

\bibitem{Beggs:2003ne}
E.~Beggs and S.~Majid, ``{Semiclassical differential structures},''
\href{http://arxiv.org/abs/math/0306273}{{\tt arXiv:math/0306273 [math-qa]}}.

\bibitem{McCurdy:2008ew}
A.~Tagliaferro, ``{The Star Product for Differential Forms on Symplectic
  Manifolds},''
\href{http://arxiv.org/abs/0809.4717}{{\tt arXiv:0809.4717 [hep-th]}}.

\bibitem{McCurdy:2009xz}
S.~McCurdy and B.~Zumino, ``{Covariant Star Product for Exterior Differential
  Forms on Symplectic Manifolds},''
  \href{http://dx.doi.org/10.1063/1.3327559}{{\em AIP Conf.Proc.} {\bf 1200}
  (2010)  204--214},
\href{http://arxiv.org/abs/0910.0459}{{\tt arXiv:0910.0459 [hep-th]}}.

\bibitem{Quillen198589}
D.~Quillen, ``Superconnections and the chern character,''
  \href{http://dx.doi.org/http://dx.doi.org/10.1016/0040-9383(85)90047-3}{{\em
  Topology} {\bf 24} (1985) no.~1, 89 -- 95}.
  \url{http://www.sciencedirect.com/science/article/pii/0040938385900473}.

\bibitem{Iazeolla:2011cb}
C.~Iazeolla and P.~Sundell, ``{Families of exact solutions to Vasiliev's 4D
  equations with spherical, cylindrical and biaxial symmetry},''
  \href{http://dx.doi.org/10.1007/JHEP12(2011)084}{{\em JHEP} {\bf 1112} (2011)
   084},
\href{http://arxiv.org/abs/1107.1217}{{\tt arXiv:1107.1217 [hep-th]}}.

\bibitem{Vasiliev:2015mka}
M.~Vasiliev, ``{Invariant Functionals in Higher-Spin Theory},''
\href{http://arxiv.org/abs/1504.07289}{{\tt arXiv:1504.07289 [hep-th]}}.

\bibitem{Didenko:2008va}
V.~E. Didenko, A.~S. Matveev, and M.~A. Vasiliev, ``{Unfolded Description of
  $AdS_4$ Kerr Black Hole},''
  \href{http://dx.doi.org/10.1016/j.physletb.2008.05.067}{{\em Phys. Lett.}
  {\bf B665} (2008)  284--293},
\href{http://arxiv.org/abs/0801.2213}{{\tt arXiv:0801.2213 [gr-qc]}}.

\bibitem{Vasiliev:2012vf}
M.~A. Vasiliev, ``{Holography, Unfolding and Higher-Spin Theory},''
  \href{http://dx.doi.org/10.1088/1751-8113/46/21/214013}{{\em J.Phys.} {\bf
  A46} (2013)  214013},
\href{http://arxiv.org/abs/1203.5554}{{\tt arXiv:1203.5554 [hep-th]}}.

\bibitem{Vasiliev:2015wma}
M.~A. Vasiliev, ``{Star-Product Functions in Higher-Spin Theory and
  Locality},'' \href{http://dx.doi.org/10.1007/JHEP06(2015)031}{{\em JHEP} {\bf
  06} (2015)  031},
\href{http://arxiv.org/abs/1502.02271}{{\tt arXiv:1502.02271 [hep-th]}}.

\bibitem{zachos2005quantum}
C.~Zachos, D.~Fairlie, and T.~Curtright, {\em Quantum mechanics in phase space:
  an overview with selected papers}, vol.~34.
\newblock World Scientific Publishing Company Incorporated, 2005.

\bibitem{Iazeolla:2007wt}
C.~Iazeolla, E.~Sezgin, and P.~Sundell, ``{Real Forms of Complex Higher Spin
  Field Equations and New Exact Solutions},''
  \href{http://dx.doi.org/10.1016/j.nuclphysb.2007.08.002}{{\em Nucl. Phys.}
  {\bf B791} (2008)  231--264},
\href{http://arxiv.org/abs/0706.2983}{{\tt arXiv:0706.2983 [hep-th]}}.

\bibitem{Vasiliev:1989re}
M.~A. Vasiliev, ``Higher spin algebras and quantization on the sphere and
  hyperboloid,'' \href{http://dx.doi.org/10.1142/S0217751X91000605}{{\em
  Int.J.Mod.Phys.} {\bf A6} (1991)  1115--1135}.

\bibitem{Bargmann:1946me}
V.~Bargmann, ``{Irreducible unitary representations of the Lorentz group},''
\href{http://dx.doi.org/10.2307/1969129}{{\em Annals Math.} {\bf 48} (1947)
  568--640}.

\bibitem{Barut:1965}
A.~O. Barut and C.~Fronsdal, ``On non-compact groups, ii. representations of
  the 2+1 lorentz group,'' {\em Proc. Roy. Soc. London} {\bf A287} (1965)
  532--548.

\bibitem{Leinaas:1977fm}
J.~Leinaas and J.~Myrheim, ``{On the theory of identical particles},''
\href{http://dx.doi.org/10.1007/BF02727953}{{\em Nuovo Cim.} {\bf B37} (1977)
  1--23}.

\bibitem{Wilczek:1982wy}
F.~Wilczek, ``{Quantum Mechanics of Fractional Spin Particles},''
\href{http://dx.doi.org/10.1103/PhysRevLett.49.957}{{\em Phys.Rev.Lett.} {\bf
  49} (1982)  957}.

\bibitem{Forte:1990hd}
S.~Forte, ``{Quantum mechanics and field theory with fractional spin and
  statistics},''
\href{http://dx.doi.org/10.1103/RevModPhys.64.193}{{\em Rev.Mod.Phys.} {\bf 64}
  (1992)  193--236}.

\bibitem{Boulanger:2013naa}
N.~Boulanger, P.~Sundell, and M.~Valenzuela, ``{Three-dimensional
  fractional-spin gravity},''
  \href{http://dx.doi.org/10.1007/JHEP02(2014)052}{{\em JHEP} {\bf 1402} (2014)
   052},
\href{http://arxiv.org/abs/1312.5700}{{\tt arXiv:1312.5700 [hep-th]}}.

\bibitem{Vasiliev:1999ba}
M.~A. Vasiliev, ``{Higher spin gauge theories: Star-product and AdS space},''
\href{http://arxiv.org/abs/hep-th/9910096}{{\tt arXiv:hep-th/9910096}}.

\bibitem{Campoleoni:2013lma}
A.~Campoleoni, T.~Prochazka, and J.~Raeymaekers, ``{A note on conical solutions
  in 3D Vasiliev theory},''
  \href{http://dx.doi.org/10.1007/JHEP05(2013)052}{{\em JHEP} {\bf 05} (2013)
  052},
\href{http://arxiv.org/abs/1303.0880}{{\tt arXiv:1303.0880 [hep-th]}}.

\bibitem{Khesin:1994ey}
B.~Khesin and F.~Malikov, ``{Universal Drinfeld-Sokolov reduction and matrices
  of complex size},'' \href{http://dx.doi.org/10.1007/BF02101626}{{\em Commun.
  Math. Phys.} {\bf 175} (1996)  113--134},
\href{http://arxiv.org/abs/hep-th/9405116}{{\tt arXiv:hep-th/9405116
  [hep-th]}}.

\bibitem{Wigner:50}
E.~P. Wigner, ``{Do the Equations of Motion Determine the Quantum Mechanical
  Commutation Relations?},''
  \href{http://dx.doi.org/10.1103/PhysRev.77.711}{{\em Phys. Rev.} {\bf 77}
  (1950)  711--712}.

\bibitem{Yang:51}
L.~M. Yang, ``{A Note on the Quantum Rule of the Harmonic Oscillator},''
  \href{http://dx.doi.org/10.1103/PhysRev.84.788}{{\em Phys. Rev.} {\bf 84}
  (1951)  788--790}.

\bibitem{Plyushchay:1994re}
M.~Plyushchay, ``{Deformed Heisenberg algebra, fractional spin fields and
  supersymmetry without fermions},''
  \href{http://dx.doi.org/10.1006/aphy.1996.0012}{{\em Annals Phys.} {\bf 245}
  (1996)  339--360},
\href{http://arxiv.org/abs/hep-th/9601116}{{\tt arXiv:hep-th/9601116
  [hep-th]}}.

\bibitem{Plyushchay:1997ty}
M.~S. Plyushchay, ``{Deformed Heisenberg algebra with reflection},''
  \href{http://dx.doi.org/10.1016/S0550-3213(97)00065-5}{{\em Nucl.Phys.} {\bf
  B491} (1997)  619--634},
\href{http://arxiv.org/abs/hep-th/9701091}{{\tt arXiv:hep-th/9701091
  [hep-th]}}.

\bibitem{Majorana:1932rj}
E.~Majorana, ``{Relativistic theory of particles with arbitrary intrinsic
  momentum},''
\href{http://dx.doi.org/10.1007/BF02959557}{{\em Nuovo Cim.} {\bf 9} (1932)
  335--344}.

\bibitem{Sudarshan:1970ss}
E.~C.~G. Sudarshan and N.~Mukunda, ``{Quantum theory of the infinite-component
  majorana field and the relation of spin and statistics},''
\href{http://dx.doi.org/10.1103/PhysRevD.1.571}{{\em Phys. Rev.} {\bf D1}
  (1970)  571--583}.

\bibitem{Dirac:1971cy}
P.~A.~M. Dirac, ``{A positive-energy relativistic wave equation},''
\href{http://dx.doi.org/10.1098/rspa.1971.0077}{{\em Proc. Roy. Soc. Lond.}
  {\bf A322} (1971)  435--445}; \href{http://dx.doi.org/10.1098/rspa.1972.0064}{{\em Proc. Roy. Soc. Lond.}
  {\bf A328} (1972)  1--7}.

\bibitem{Staunton:1974dy}
L.~P. Staunton, ``{A Spin 1/2 Positive Energy Relativistic Wave Equation},''
\href{http://dx.doi.org/10.1103/PhysRevD.10.1760}{{\em Phys. Rev.} {\bf D10}
  (1974)  1760}.

\bibitem{Vasiliev:1992gr}
M.~A. Vasiliev, ``{Unfolded representation for relativistic equations in (2+1)
  anti-De Sitter space},''
{\em Class. Quant. Grav.} {\bf 11} (1994)  649--664.

\bibitem{Vasiliev:1992ix}
M.~A. Vasiliev, ``{Equations of motion for d = 3 massless fields interacting
  through Chern-Simons higher spin gauge fields},''
\href{http://dx.doi.org/10.1142/S0217732392003116}{{\em Mod. Phys. Lett.} {\bf
  A7} (1992)  3689--3702}.

\bibitem{Kessel:2015kna}
P.~Kessel, G.~L. Gomez, E.~Skvortsov, and M.~Taronna, ``{Higher Spins and
  Matter Interacting in Dimension Three},''
\href{http://arxiv.org/abs/1505.05887}{{\tt arXiv:1505.05887 [hep-th]}}.

\bibitem{Iazeolla:2015tca}
C.~Iazeolla and J.~Raeymaekers, ``{On big crunch solutions in
  Prokushkin-Vasiliev theory},''
  \href{http://dx.doi.org/10.1007/JHEP01(2016)177}{{\em JHEP} {\bf 01} (2016)
  177},
\href{http://arxiv.org/abs/1510.08835}{{\tt arXiv:1510.08835 [hep-th]}}.

\bibitem{Vasiliev:2007yc}
M.~Vasiliev, ``{On Conformal, SL(4,R) and Sp(8,R) Symmetries of 4d Massless
  Fields},'' \href{http://dx.doi.org/10.1016/j.nuclphysb.2007.10.017}{{\em
  Nucl.Phys.} {\bf B793} (2008)  469--526},
  \href{http://arxiv.org/abs/0707.1085}{{\tt arXiv:0707.1085 [hep-th]}}.

\bibitem{Sezgin:2012ag}
E.~Sezgin and P.~Sundell, ``{Supersymmetric Higher Spin Theories},''
  \href{http://dx.doi.org/10.1088/1751-8113/46/21/214022}{{\em J.Phys.} {\bf
  A46} (2013)  214022},
\href{http://arxiv.org/abs/1208.6019}{{\tt arXiv:1208.6019 [hep-th]}}.

\bibitem{Prokushkin:1999gc}
S.~F. Prokushkin, A.~Y. Segal, and M.~A. Vasiliev, ``{Coordinate free action
  for AdS(3) higher spin matter systems},''
  \href{http://dx.doi.org/10.1016/S0370-2693(00)00258-6}{{\em Phys.Lett.} {\bf
  B478} (2000)  333--342},
\href{http://arxiv.org/abs/hep-th/9912280}{{\tt arXiv:hep-th/9912280
  [hep-th]}}.

\bibitem{DeWildeLecomte}
M.~De~Wilde and P.~B. Lecomte, ``Existence of star-products and of formal
  deformations of the poisson lie algebra of arbitrary symplectic manifolds,''
  {\em Letters in Mathematical Physics} {\bf 7} (1983) no.~6, 487--496.

\bibitem{Fedosov:1994zz}
B.~v. Fedosov, ``{A Simple geometrical construction of deformation
  quantization},''
{\em J. Diff. Geom.} {\bf 40} (1994) no.~2, 213--238.

\bibitem{Cattaneo:1999fm}
A.~S. Cattaneo and G.~Felder, ``{A path integral approach to the Kontsevich
  quantization formula},'' \href{http://dx.doi.org/10.1007/s002200000229}{{\em
  Commun. Math. Phys.} {\bf 212} (2000)  591--611},
\href{http://arxiv.org/abs/math/9902090}{{\tt arXiv:math/9902090}}.

\bibitem{Cattaneo:2001vu}
A.~S. Cattaneo and G.~Felder, ``{On the globalization of Kontsevich's star
  product and the perturbative Poisson sigma model},''
  \href{http://dx.doi.org/10.1143/PTPS.144.38}{{\em Prog. Theor. Phys. Suppl.}
  {\bf 144} (2001)  38--53},
\href{http://arxiv.org/abs/hep-th/0111028}{{\tt arXiv:hep-th/0111028
  [hep-th]}}.

\end{thebibliography}

\end{document}